%% file: EXO-21-007_temp.tex
\pdfoutput=1
\documentclass[11pt,twoside,a4paper,cmspaper,final,collab]{cms-tdr}
\def\svnVersion{3dd28d4}\def\svnDate{2024/07/16}\def\cmsCernNoTag{CERN-EP-2023-143}\def\cmsCernDate{\today}\def\cmsMessage{Published in Physical Review D as \href{http://dx.doi.org/10.1103/PhysRevD.110.012010}{\doi{10.1103/PhysRevD.110.012010}.}}
\begin{document}\cmsNoteHeader{EXO-21-007}

\renewcommand{\cmsCollabName}{The CMS and TOTEM Collaborations}
\renewcommand{\cmsTag}{CMS-\cmsNUMBER\\&TOTEM 2023-003\\}
\renewcommand{\cmsPubBlock}{\begin{tabular}[t]{@{}r@{}l}&CMS \cmsSTYLE~\cmsNUMBER\\&TOTEM 2023-003\\\end{tabular}}
\renewcommand{\cmsCopyright}{\copyright\,\the\year\ CERN for the benefit of the CMS and TOTEM Collaborations.}
\renewcommand{\appMsg}{See Appendices~\ref{app:collab} and~\ref{app:totem} for the lists of collaboration members}
\renewcommand{\cmslogo}{\includegraphics[height=2cm]{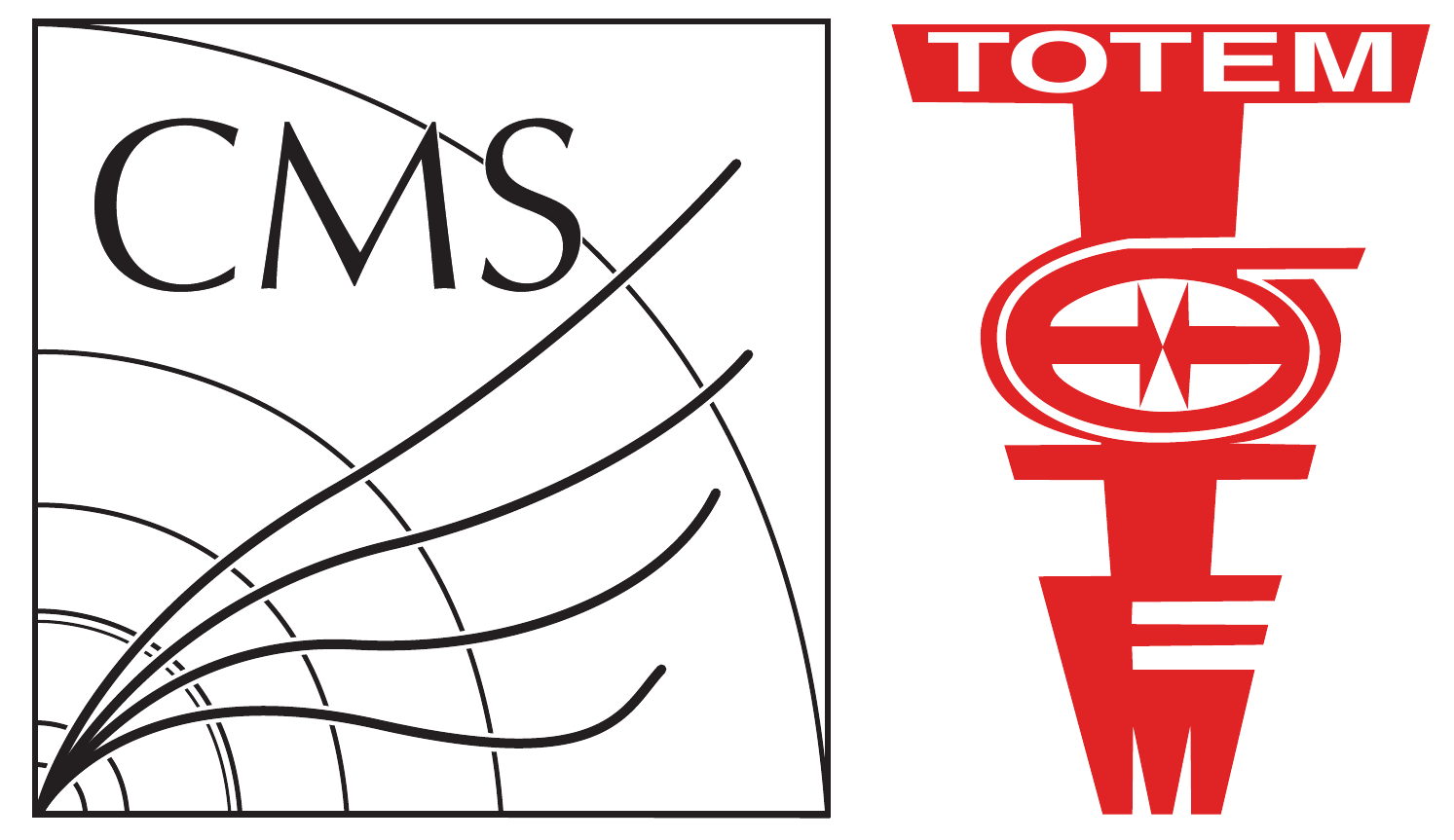}}

\newlength\cmsTabSkip\setlength{\cmsTabSkip}{1ex}

\cmsNoteHeader{EXO-21-007}

\title{Search for high-mass exclusive diphoton production with tagged protons in proton-proton collisions at \texorpdfstring{$\sqrt{s} = 13\TeV$}{sqrt(s) = 13 TeV}}

\author*[cern]{The CMS and TOTEM Collaborations}

\date{\today}

\abstract{
A search is presented for high-mass exclusive diphoton production via photon-photon fusion in proton-proton collisions at $\sqrt{s} = 13\TeV$ in events where both protons survive the interaction. The analysis utilizes data corresponding to an integrated luminosity of 103\fbinv collected in 2016--2018 with the central CMS detector and the CMS and TOTEM precision proton spectrometer (PPS). Events that have two photons with high transverse momenta ($\pt^{\PGg} > 100\GeV$), back-to-back in azimuth, and with a large diphoton invariant mass ($m_{\PGg\PGg} > 350\GeV$) are selected. To remove the dominant inclusive diphoton backgrounds, the kinematic properties of the protons detected in PPS are required to match those of the central diphoton system. Only events having opposite-side forward protons detected with a fractional momentum loss between 0.035 and 0.15 (0.18) for the detectors on the negative (positive) side of CMS are considered. One exclusive diphoton candidate is observed for an expected background of 1.1 events. Limits at 95\% confidence level are derived for the four-photon anomalous coupling parameters $\abs{\zeta_1} < 0.073\TeV^{-4}$ and $\abs{\zeta_2} < 0.15\TeV^{-4}$, using an effective field theory. Additionally, upper limits are placed on the production of axion-like particles with coupling strength to photons $f^{-1}$ that varies from $0.03\TeV^{-1}$ to $1\TeV^{-1}$ over the mass range from 500 to 2000\GeV.
}

\hypersetup{%
pdfauthor={CMS Collaboration},%
pdftitle={Search for high-mass exclusive diphoton production with tagged protons in proton-proton collisions at sqrt(s) = 13 TeV},%
pdfsubject={CMS},%
pdfkeywords={CMS, TTOEM, forward physics, diphotons}}

\maketitle 

\section{Introduction}

\begin{figure}[htpb!]
    \centering
    \includegraphics[width=0.4\textwidth]{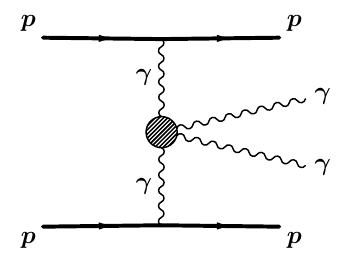}
    \caption{
    Diagram of diphoton production via photon fusion with intact protons in the final state. 
    The four-photon vertex includes virtual contributions from SM or BSM charged fermions or bosons. 
    In other BSM scenarios, a new heavy particle can be produced in the $s$-channel, such as an axion-like particle that decays into two photons.
}
    \label{fig:diag_gggg}
\end{figure}

While the standard model (SM) of particle physics has been largely successful in describing the known elementary particles and interactions, there are still many observed phenomena that suggest it is incomplete. 
These include the unknown nature of dark matter, the origin of matter-antimatter asymmetry, the generation of neutrino masses, and the absence of charge-parity (CP) symmetry violation in the theory of the strong interaction. 
Photon-photon interactions in proton-proton ($\Pp\Pp$) collisions at the LHC provide an environment with very small irreducible backgrounds to probe theories that go beyond the standard model (BSM) and that feature anomalous quartic gauge couplings (aQGC) at energies so-far unexplored. 
At the LHC, when the protons interact purely electromagnetically the final state can be particularly simple with a C-even, neutral system produced exclusively in the central detector and the two protons remaining intact (Fig.~\ref{fig:diag_gggg}). 
A process of interest is that of the exclusive production of two photons in photon fusion, usually called light-by-light (LbL) scattering~\cite{dEnterria:2013zqi}, which allows the probing of multiple BSM phenomena including the production of axion-like particles (ALPs)~\cite{Knapen:2016moh,Baldenegro:2018hng}, monopoles~\cite{Kurochkin:2006jr}, and massive gravitons~\cite{Atwood:1999zg,Inan:2012zz,dEnterria:2023npy}, as well as searches for Born--Infeld nonlinear extensions of quantum electrodynamics~\cite{Ellis:2017edi}, extra spatial dimensions~\cite{Lietti:2002rq}, etc.

Both the CMS and ATLAS collaborations have presented evidence or observation of the SM LbL process in electromagnetic ultraperipheral collisions (UPCs) of  heavy ions~\cite{Sirunyan:2018fhl,Aad:2019ock}. 
However, UPCs with ions cannot reach two-photon invariant masses ($m_{\PGg\PGg}$) much above 100\GeV~\cite{Baltz:2007kq,Bruce:2018yzs}, where larger contributions to the four-photon ($4\PGg$) interaction from BSM physics can be expected. 
Proton-proton collisions, on the other hand, feature larger beam energies and luminosities, and harder photon spectra than heavy ion UPCs. 
By tagging the intact final-state protons, one can better identify the process and study $4\PGg$ anomalous couplings at the multi-\TeVns scale~\cite{Fichet:2014uka}.
Different particles can contribute to the four-photon interaction shown in Fig.~\ref{fig:diag_gggg}. This can include loops of SM or of new charged fermions or bosons. In other BSM scenarios, a C-even scalar, pseudoscalar, or tensor neutral resonance can be produced that decays back to two photons.

As shown in Ref.~\cite{Fichet:2014uka}, with the assumption of a new mass scale heavier than the current experimentally reachable energy, the anomalous $4\PGg$ interactions can be described by an effective Lagrangian density using dimension-8 operators,
\begin{equation}
    \label{eqn:L_eff}
    \mathcal{L}_{4\PGg} = \zeta_1F_{\mu\nu}F^{\mu\nu}F_{\rho\sigma}F^{\rho\sigma}+\zeta_2F_{\mu\nu}F^{\nu\rho}F_{\rho\lambda}F^{\lambda\mu},
\end{equation}
where $F$ is the electromagnetic field tensor.
These $\zeta_1$ and $\zeta_2$ parameters are identically zero in the SM, but this would not be so if BSM particles contribute to the $4\PGg$ process. In this latter case, from Eq.~(\ref{eqn:L_eff}), the differential cross section for the anomalous four-photon production can be written as 
\begin{equation*}
    \label{eqn:angular_xs}
    \frac{\rd\sigma}{\rd\Omega} = \frac{1}{16\pi^2s}\left(s^2+t^2+st\right)^2 \left[48 \zeta_1^2 + 40\zeta_1\zeta_2 + 11\zeta_2^2\right],
\end{equation*}
where $s$ and $t$ are the Mandelstam variables.

The $s$-channel production of a pseudoscalar particle, such as an ALP, can also give rise to an increased production rate of the LbL process with respect to the SM expectations~\cite{Baldenegro:2018hng}. The production of ALPs is often characterized by two parameters: the ALP mass and its photon coupling, $f^{-1}\equiv C_{\PGg\PGg}/\Lambda$~\cite{Bauer:2017ris}, where $C_{\PGg\PGg}$ is a dimensionless coefficient between zero and one, and $\Lambda$ is a high-energy scale associated with the spontaneous breaking of an approximate Peccei--Quinn global U$(1)$ symmetry~\cite{Peccei:1977hh}. Our analysis considers ALPs that couple only to photons in a photophilic scenario with $C_{\PGg\PGg} = 1$, as often considered in the literature~\cite{Agrawal:2021dbo}.

The strategy of this study is to search for central diphoton events exhibiting kinematic properties consistent with their exclusive production through photon fusion, in coincidence with two opposite-side protons having a momentum loss compatible with the rapidity of the diphoton system. Compared to the previous exclusive diphoton result reported by the CMS and TOTEM Collaborations~\cite{CMS:2021plq}, the integrated luminosity has been increased by more than a factor of ten, and a set of new techniques are used to improve the sensitivity to the $4\PGg$ process.
The new techniques include a more efficient photon identification algorithm, improved forward proton reconstruction, and a more robust background estimation. Tabulated results are provided in the HEPData record for this analysis~\cite{hepdata}.

\section{The CMS detector and event selection and reconstruction}
\label{sec:detector}

This analysis relies equally on the central CMS and Precision Proton Spectrometer (PPS) detectors shown in Fig.~\ref{fig:pps_schematic}.
Both of these are described below along with the event reconstruction algorithms.

\subsection{The central CMS detector}

The main feature of the CMS apparatus is a superconducting solenoid of 6\unit{m} internal diameter, providing a magnetic field of 3.8\unit{T}. Within the solenoid volume are a silicon pixel and strip tracker, a lead tungstate crystal electromagnetic calorimeter (ECAL), and a brass and scintillator hadron calorimeter (HCAL), each composed of a barrel and two endcap sections. Forward calorimeters extend the pseudorapidity coverage provided by the barrel and endcap detectors. Muons are detected in gas-ionization chambers embedded in the steel flux-return yoke outside the solenoid. 

The electromagnetic calorimeter consists of 75\,848 lead tungstate crystals, which provide coverage in pseudorapidity $\abs{\eta} < 1.48 $ and $1.48 < \abs{\eta} < 3.0$ in a barrel region (EB) and two endcap regions (EE), respectively. 
Preshower detectors consisting of two planes of silicon sensors interleaved with a total of three radiation lengths of lead are located in front of each EE detector. 

In the EB, an energy resolution of about 1\% is achieved for unconverted or late-converting photons in the tens of \GeVns energy range. 
The remaining barrel photons have a resolution of about 1.3\% extending up to a pseudorapidity of $\abs{\eta} = 1$, rising to about 2.5\% at $\abs{\eta} = 1.4$. 
In the endcaps, the resolution of unconverted or late-converting photons is about 2.5\%, while the remaining endcap photons have a resolution between 3 and 4\%~\cite{CMS:EGM-14-001}. 

A more detailed description of the CMS detector, together with a definition of the coordinate system used and the relevant kinematic variables, can be found in~\cite{Chatrchyan:2008zzk}.

Events of interest are selected using a two-tiered trigger system. The first level, composed of custom hardware processors, uses information from the calorimeters and muon detectors to select events at a rate of around 100\unit{kHz} within a fixed latency of about 4\mus~\cite{Sirunyan:2020zal}. The second level, known as the high-level trigger (HLT), consists of a farm of processors running a version of the full event reconstruction software optimized for fast processing, and reduces the event rate to around 1\unit{kHz} before data storage~\cite{Khachatryan:2016bia}. 

The global event reconstruction, also called particle-flow (PF) event reconstruction~\cite{CMS-PRF-14-001}, aims to reconstruct and identify each individual particle in an event, with an optimized combination of all subdetector information. In this process, the identification of the particle type (photon, electron, muon, charged or neutral hadron) plays an important role in the determination of the particle direction and energy. Photons are identified as ECAL energy clusters not linked to the extrapolation of any charged particle trajectory to the ECAL. 
Electrons are identified by a primary charged particle track and potentially many ECAL energy clusters corresponding to the extrapolation of this track to the ECAL as well as possible bremsstrahlung photons emitted along the way through the tracker material.
Muons (\eg, from semileptonic decays of heavy-quark hadrons) are identified as tracks in the central tracker consistent with either a track or several hits in the muon system, and associated with calorimeter deposits compatible with the muon hypothesis. Charged hadrons are identified as charged particle tracks that are not identified as electrons, nor as muons. Finally, neutral hadrons are identified as HCAL energy clusters not linked to any charged hadron trajectory, or as a combined ECAL and HCAL energy excess with respect to the expected charged hadron energy deposit.

The energy of photons is obtained from the ECAL measurement. The energy of electrons is determined from a combination of the track momentum at the main interaction vertex, the corresponding ECAL cluster energy, and the energy sum of all bremsstrahlung photons associated with the track. The energy of muons is obtained from the corresponding track momentum. The energy of charged hadrons is determined from a combination of the track momentum and the corresponding ECAL and HCAL energies, corrected for the response function of the calorimeters to hadronic showers. Finally, the energy of neutral hadrons is obtained from the corresponding corrected ECAL and HCAL energies.

For each event, hadronic jets are clustered from these reconstructed particles using the infrared- and collinear-safe anti-\kt algorithm~\cite{Cacciari:2008gp, Cacciari:2011ma} with a distance parameter of 0.4. Additional $\Pp\Pp$ interactions within the same or nearby bunch crossings (pileup) can contribute additional tracks and calorimetric energy depositions. To mitigate this effect, charged particles identified to be originating from pileup vertices are discarded and an offset correction is applied to correct for remaining contributions~\cite{CMS:2020ebo}.

\subsection{The Precision Proton Spectrometer}
\label{sec:pps}

\begin{figure*}[htpb!]
    \centering
    \includegraphics[width=\textwidth]{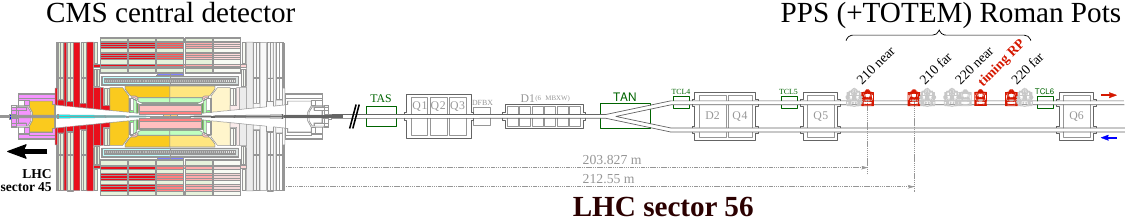}
    \caption{A schematic view of one side of the PPS detector with respect to the central CMS detector.
    The stations labeled as (or located at) 210 and 220\unit{m} house the Roman Pot detectors.
    Timing detector stations are also shown, although they are not used in this analysis.
    A symmetric set of detectors exists on the opposite side of CMS as well.}
    \label{fig:pps_schematic}
\end{figure*}

The PPS detectors were developed by a collaboration between the CMS and TOTEM experiments (previously named CT-PPS)~\cite{CMS:2014sdw}. These detectors extend the CMS physics program to study central-exclusive production processes~\cite{LHCForwardPhysicsWorkingGroup:2016ote} under high-luminosity conditions at the LHC, in the context of both SM and BSM physics studies.
The PPS detectors are located approximately 210 and 220\unit{m} away from the LHC interaction point 5 (IP5) on both sides of CMS, as shown in Fig.~\ref{fig:pps_schematic} for one side.

After a $\Pp\Pp$ collision at IP5, if a proton remains intact in the final state, it may continue to travel down the LHC beam line, guided by the LHC magnets, until it interacts with the PPS detectors in the forward regions.
The horizontal distance of the proton from the center of the beam is proportional to the momentum loss of the proton, and in this way the LHC magnetic lattice acts as a mass spectrometer for measuring protons.
The main observable used in the analysis is the fractional momentum loss of the proton, denoted as $\xi$.
When measuring each of the final-state protons from an exclusive event, their $\xi_\Pp$ value can be used to derive the mass and rapidity of the centrally produced system at the $\Pp\Pp$ center-of-mass energy $\sqrt{s}$, as follows:
\begin{equation}
\label{eqn:massrap}
  m_{\Pp\Pp}=\sqrt{s\xi^+_\Pp \xi^-_\Pp},\hspace{1em} y_{\Pp\Pp}=\frac{1}{2}\log{\left(\frac{\xi^+_\Pp}{\xi^-_\Pp}\right)}.
\end{equation}
The positive and negative indices of $\xi_\Pp$ refer to protons moving in the positive and negative $z$ directions, respectively.
The PPS detectors located in the positive-$z$ direction from CMS are referred to as sector-45 (for the region between IP4 and IP5), and the PPS detectors located in the negative-$z$ direction from CMS are referred to as sector-56 (for the region between IP5 and IP6) as shown in Fig.~\ref{fig:pps_schematic}.
At each of the 210 and 220\unit{m} stations, movable, near-beam insertions called Roman Pots (RPs) provide the housing for the proton tracking detectors. The PPS tracking detector configuration was different for each of the three data-taking years: in 2016 with silicon microstrip detectors in both tracking Roman pots in the same arm, in 2017 with silicon microstrip detectors in the one and silicon pixel detectors in the other Roman pot in the same arm, and finally in 2018 with silicon pixel detectors in both. The protons are reconstructed using a ``multi-RP'' algorithm, which combines tracks reconstructed in both of the tracking Roman pots in each arm of PPS. A detailed description of the PPS detector system, alignment, optics, and simulation can be found in~\cite{TOTEM:2022vox}.

During 2016--2018, PPS collected proton-proton collision data corresponding to an integrated luminosity of 107.7\fbinv.
Using data from 2016, corresponding to an integrated luminosity of 9.4\fbinv, the first validation of the PPS system resulted in the observation of the SM exclusive production of a pair of charged leptons, $\PGg\PGg\to\Pell^+\Pell^-$~\cite{CMS:2018uvs}.
Signal events were selected by requiring a match between the dilepton and forward protons kinematic properties.
This study revealed a signal yield corresponding to a significance greater than 5.1 standard deviations above the background-only expectation, and constituted the first observation of this process at masses around and above the electroweak scale.

\section{Data sets and simulated samples}
\label{sec:samples}

This analysis uses data sets collected by the CMS and TOTEM experiments in 2016, 2017, 2018 of $\Pp\Pp$ collisions at $\sqrt{s}$ = 13\TeV corresponding to integrated luminosities of 9.8~\cite{CMS:2021xjt}, 37.2~\cite{CMS:LUM-17-004}, and 55.7~\cite{CMS:LUM-18-002}\fbinv, respectively. The data sets comprise the LHC running periods where all central CMS and PPS detectors were operating simultaneously and included in the data acquisition.

Simulated Monte Carlo (MC) samples are used to model the signal and SM backgrounds. The dominant background is due to inclusive $\PGg \PGg + \text{jet(s)}$ production, where ``jet(s)'' is loosely used to indicate any additional energy in the event. Subleading backgrounds include $\ttbar+ \text{jet(s)}$ and $\PV + \PGg$ processes, where \PV represents a weak boson. 
The decay channels considered for weak bosons are $\PZ\to\Pell^+\Pell^-$ and $\PWpm\to\Pell^\pm \Pgn_{\Pell}$. Background samples are modeled with the \textsc{MadGraph5\_aMC@NLO}~\cite{Alwall:2014hca} package, with the 2016 samples generated using the \textsc{NNPDF3.0} parton distribution functions (PDFs) at next-to-leading order (NLO) accuracy, and the other samples with \textsc{NNPDF3.1} PDFs at next-to-NLO accuracy. To estimate additional backgrounds from quantum chromodynamics (QCD) jet production, an electron- and photon-enriched QCD sample generated with \PYTHIA 8.2~\cite{Sjostrand:2014zea} and the CP5 underlying event tune~\cite{Sirunyan:2019dfx} is utilized. 

\begin{figure*}[htpb!]
    \centering
    \includegraphics[width=0.49\textwidth]{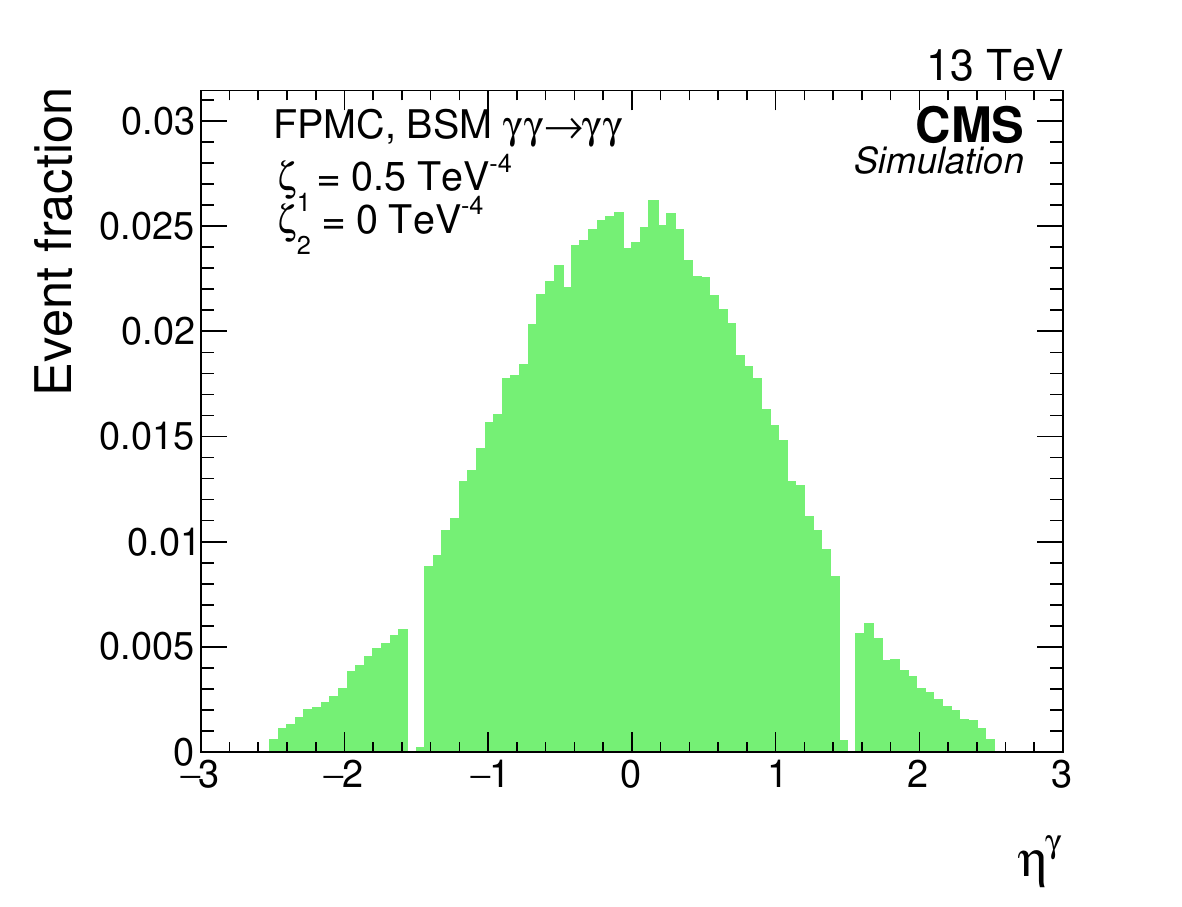}
    \includegraphics[width=0.49\textwidth]{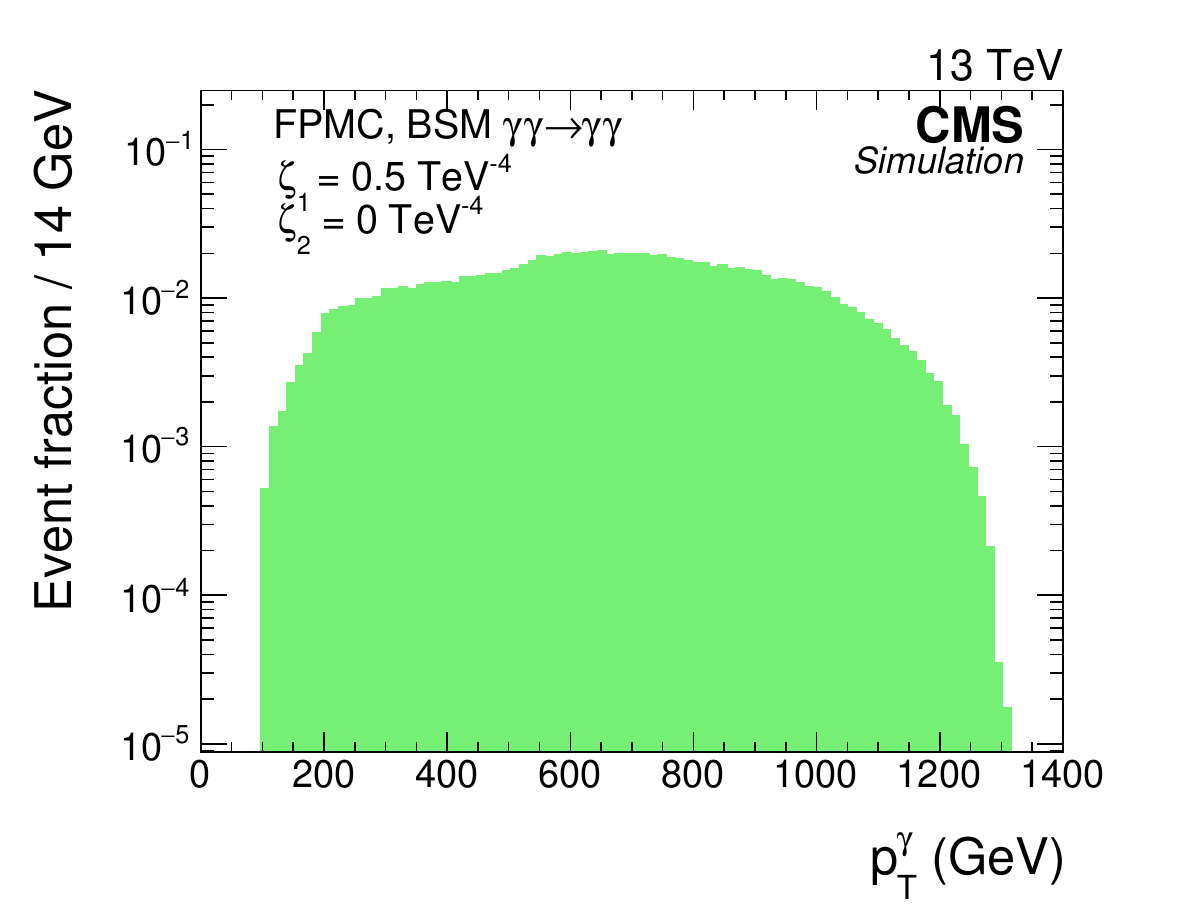}
    \includegraphics[width=0.49\textwidth]{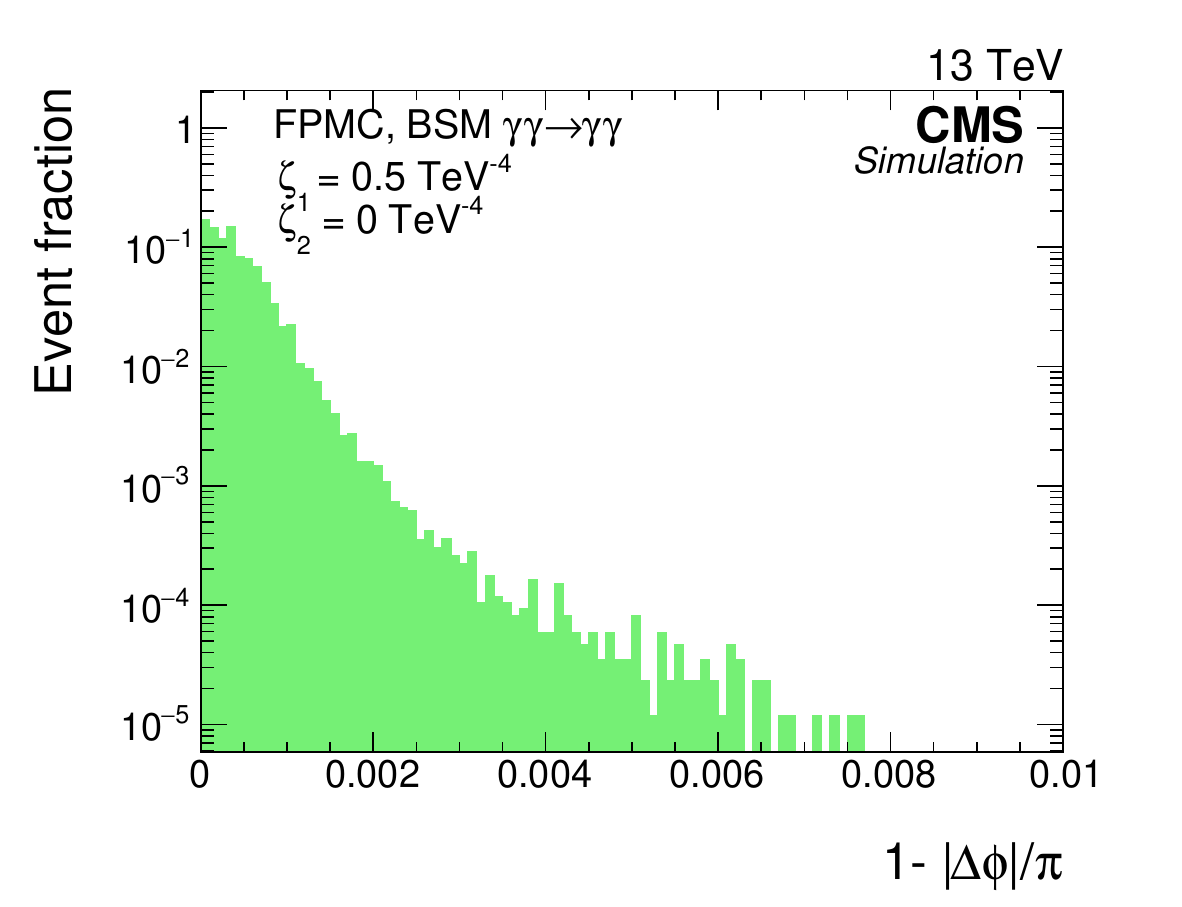}
    \includegraphics[width=0.49\textwidth]{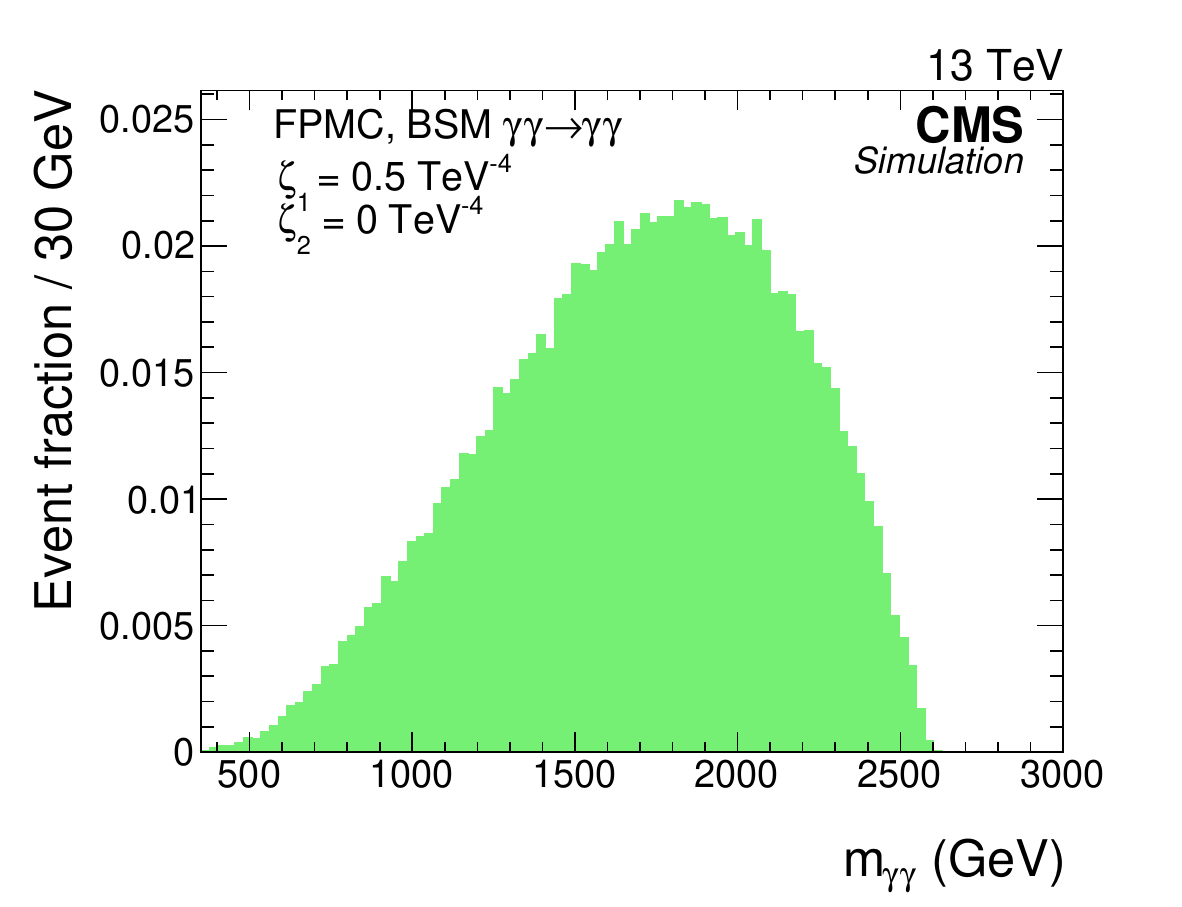}
    \caption{Kinematic distributions of simulated aQGC $\PGg\PGg \to \PGg\PGg$ events: Single-photon $\eta$ (upper left) and \PT (upper right), and diphoton acoplanarity (lower left) and mass (lower right). The events are generated with FPMC, reconstructed accounting for the full CMS detector response, for aQGC parameters $\zeta_1 = 0.5 \TeV^{-4}$ and $\zeta_2 = 0 \TeV^{-4}$. A preselection is applied to these events as described in Section~\ref{sec:selection}.}
    \label{fig:kinematics_signal}
\end{figure*}

\begin{figure*}[htpb!]
    \centering
    \includegraphics[width=0.49\textwidth]{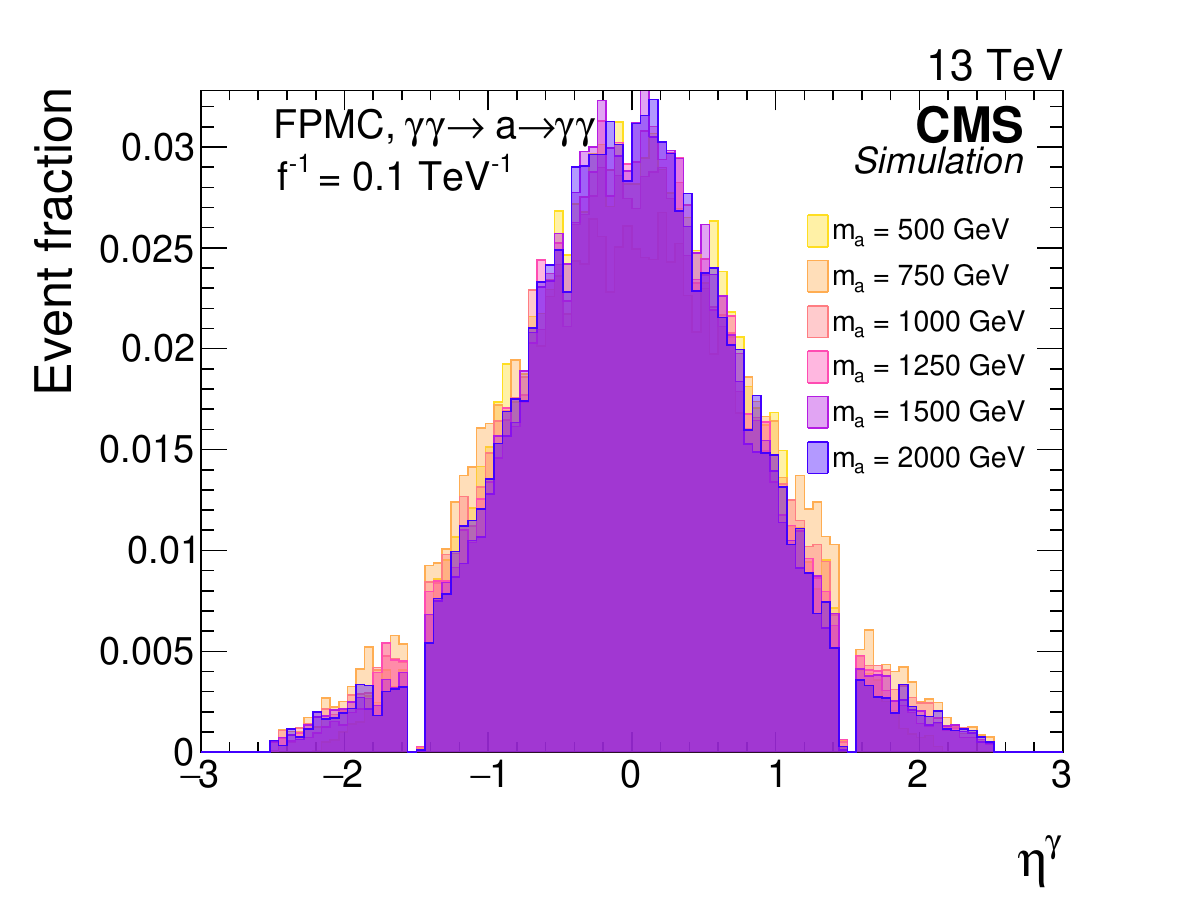}
    \includegraphics[width=0.49\textwidth]{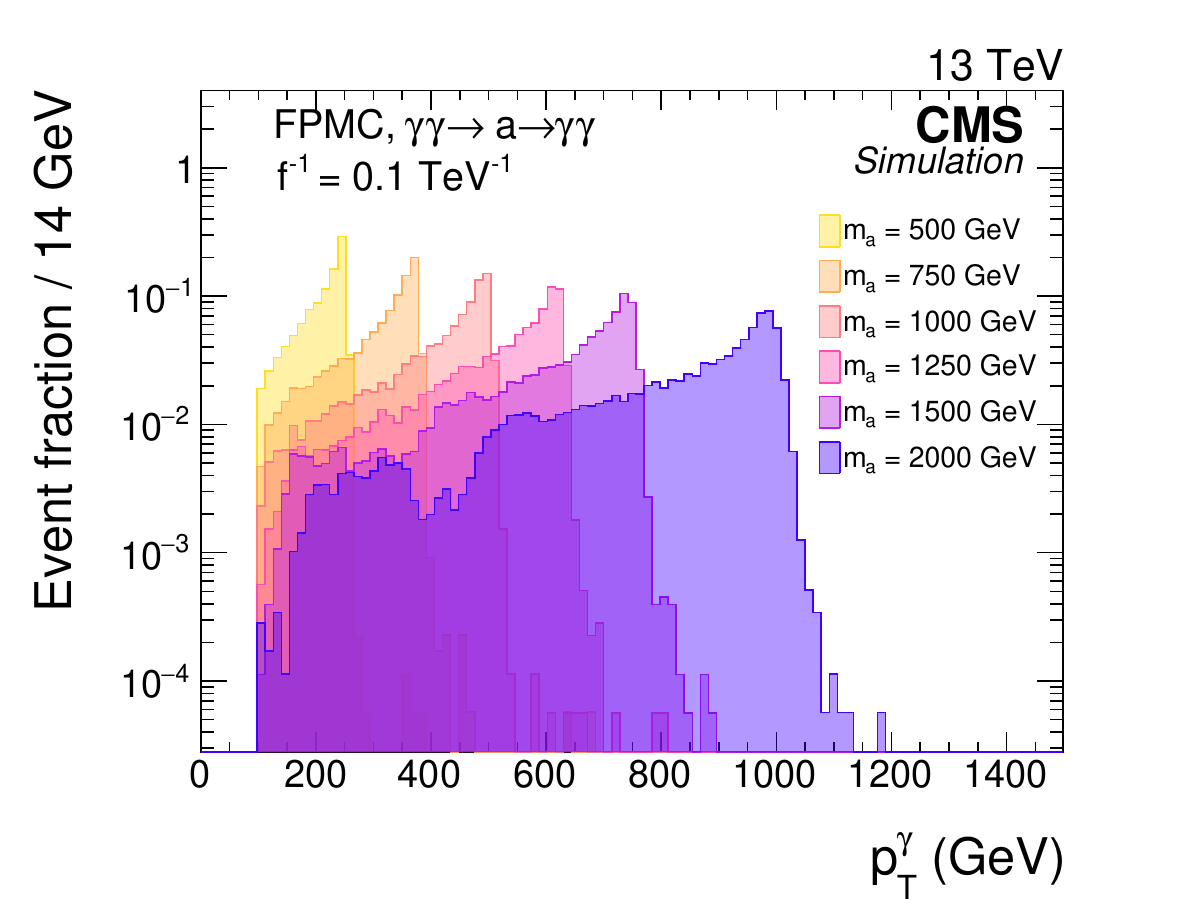}
    \includegraphics[width=0.49\textwidth]{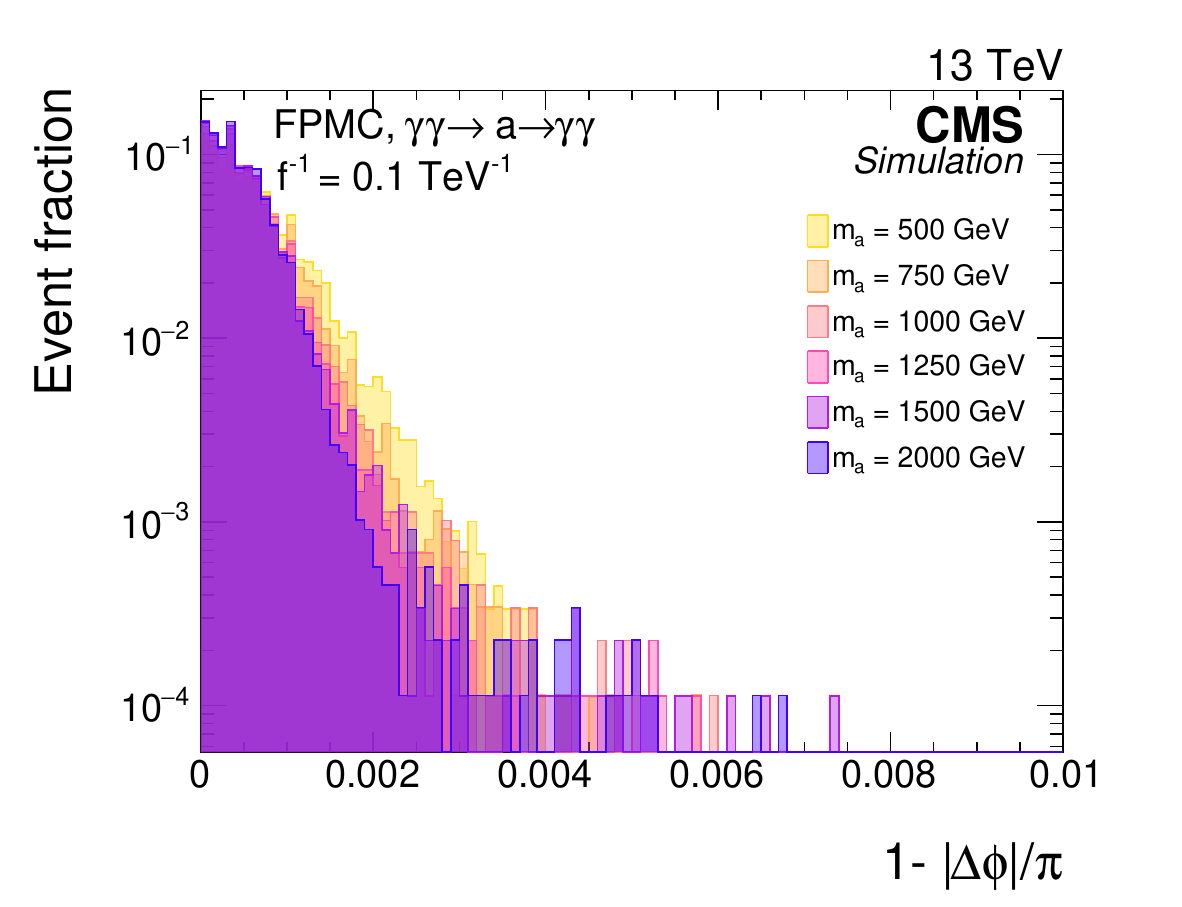}
    \includegraphics[width=0.49\textwidth]{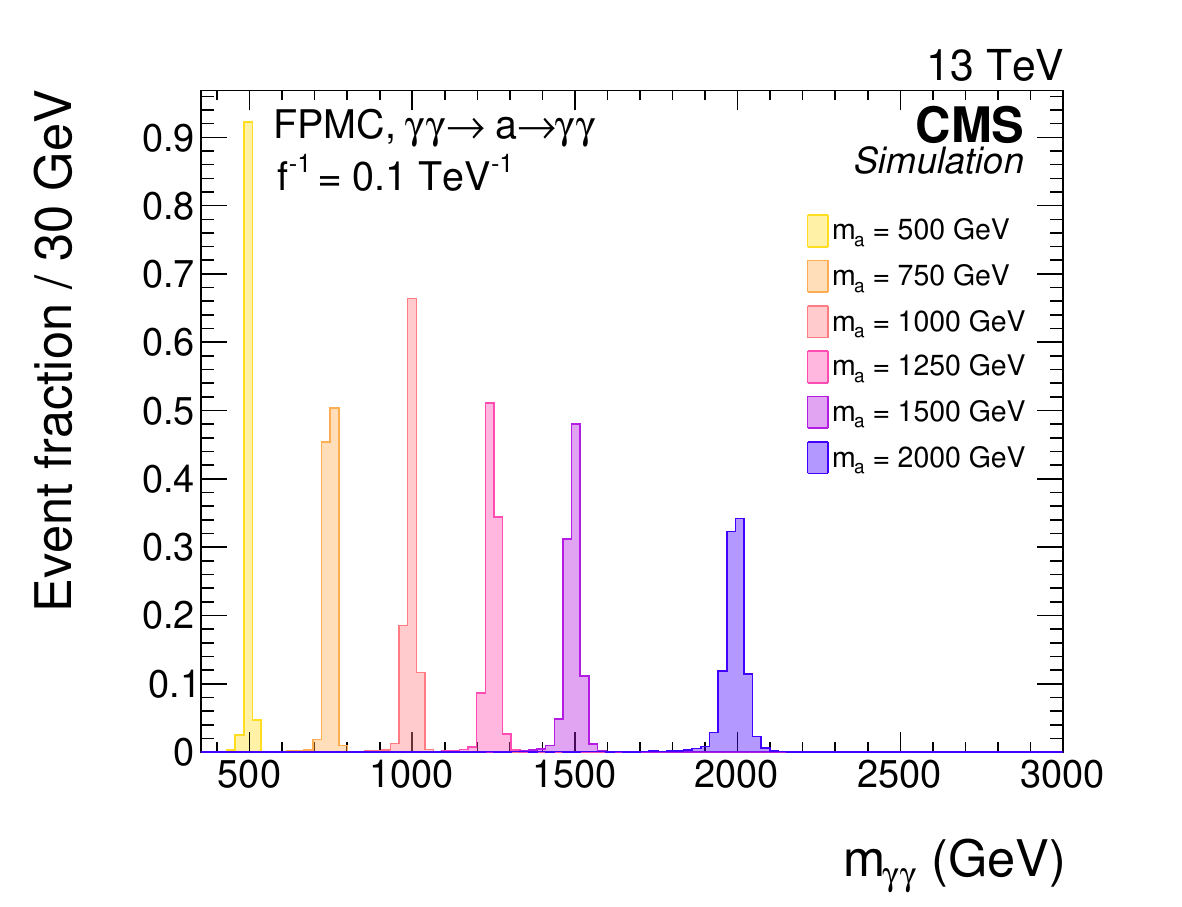}
    \caption{Kinematic distributions of simulated ALP $\PGg\PGg \to a \to \PGg\PGg$ events: Single-photon $\eta$ (upper left) and \PT (upper right), and diphoton acoplanarity (lower left) and mass (lower right). The events are generated with FPMC, reconstructed accounting for the full CMS detector response, for a value $f^{-1} =0.1\TeV^{-1}$ of the ALP-photon coupling. A preselection is applied to these events as described in Section~\ref{sec:selection}.}
    \label{fig:kinematics_alp}
\end{figure*}

The LbL signal samples are generated using the Forward Physics Monte Carlo (FPMC) code~\cite{Boonekamp:2011ky} assuming collinear photon emission from both incoming protons. The photon fluxes from each proton are modelled using the Budnev \textit{et al.} parameterization~\cite{Budnev:1974de}. For the simulated conditions of each data-taking year, various $\PGg\PGg \to \PGg\PGg$ aQGC signal samples are generated for representative $\zeta_1$ and $\zeta_2$ values. Kinematic distributions of the simulated aQGC events are shown in Fig.~\ref{fig:kinematics_signal}. Additionally, samples modeling ALP production via $s$-channel exchange, $\PGg\PGg \to a \to \PGg\PGg$, are produced for twelve different ALP masses in the range of 500--2000\GeV where the analysis reaches the best sensitivity. These simulated data samples are used to derive the corresponding reconstruction efficiencies.
Kinematic distributions for various ALP mass samples are shown in Fig.~\ref{fig:kinematics_alp}. The shape of these distributions are independent of the coupling strength, $f^{-1}$~\cite{Baldenegro:2018hng}. It should be noted that the LbL, aQGC, and ALP processes at any given photon-photon mass value all have similar kinematic properties, and are studied using the same event selection criteria.

The parton showering, hadronization, and particle decays in the simulated background samples are carried out with \PYTHIA~8.230 for the 2016 and 2018 samples, and 8.226 for the 2017 samples. All generated samples are subsequently passed through the \GEANTfour simulated response of the CMS detector~\cite{Agostinelli:2002hh}.
In all cases, the pileup distribution in the simulated events is weighted to match that observed in data.

\section{Photon identification and isolation}
\label{sec:id}

A multivariate analysis (MVA) utilizing a boosted decision tree (BDT) is used for identification (ID) and isolation of photons.
This technique allows for the definition of a single discriminating variable, based on many parameters for each photon candidate, that helps separate prompt isolated photons from backgrounds.
The inputs to the BDT are shower shape and isolation variables, as well as quantities used to minimize the pileup present in the event~\cite{CMS:EGM-14-001}, described next.
The variables used by the BDT are described in Table \ref{tab:bdt}.

\begin{table}[htb]
   \centering
    \topcaption{The input variables to the BDT used in the photon identification process.}
    \begin{scotch}{ll}
        BDT variable        & Description\\
        \hline
        $\sigma_{i\eta i\eta}$ & \begin{tabular}[t]{l}Second-order spatial moment of the\\ photon candidate with respect to $\eta$.\end{tabular}\\
        $R_9$ and $R_9(5\times 5)$       & \begin{tabular}[t]{l}Sum of the energy contained in the\\ $3\times3$ matrix centered on the most\\ energetic crystal of the supercluster\\ (SC) divided, respectively, by the SC\\ energy or by the energy contained\\ in a $5\times5$ matrix.\end{tabular}\\
        $q_{\eta\phi}$        & \begin{tabular}[t]{l}Elements of the covariance matrix\\ of the single-crystal energy values\\ in $\eta$ and $\phi$ for the $5\times5$ array of\\ crystals centered on the highest\\ energy crystal.\end{tabular}\\
        $S_4$   & \begin{tabular}[t]{l}Energy contained in the most\\ energetic $2\times2$ array of crystals\\ (containing the seed crystal) divided\\ by the SC energy.\end{tabular}\\
        $H/E$   & \begin{tabular}[t]{l}Ratio between the energy deposits\\ in the hadronic and electromagnetic\\ calorimeters.\end{tabular}\\
        $w_\eta$   & \begin{tabular}[t]{l}$\eta$ width of the electromagnetic shower.\end{tabular}\\
        $w_\phi$   & \begin{tabular}[t]{l}$\phi$ width of the electromagnetic shower.\end{tabular}\\
    \end{scotch}
    \label{tab:bdt}
\end{table}

The photon isolation variables that are considered, based on the PF algorithm, are $I_{\PGg}$, $I_\text{ch}$, and $I_\text{n}$. These are obtained by summing the transverse momenta of the photons, charged hadrons, and neutral hadrons, respectively, inside an isolation region of radius $\Delta R = \sqrt{\smash[b]{\Delta\eta\,^2 + \Delta\phi\,^2}}$ around the photon candidate~\cite{Sirunyan:2020ycc}.
Other variables used as inputs are the raw SC energy and $\eta$ of the photon candidate as well as the average energy density of the event.
Specifically for photon candidates in the endcap, the preshower energy spread in the $\eta$ direction and the preshower energy divided by the SC energy are provided to the decision tree.

The BDT is trained on simulated $\PGg + \text{jet(s)}$ events where the photon candidates have passed the set of preselection criteria listed in Table~\ref{tab:ID_preselection}. The BDT is also validated with simulated Drell--Yan dilepton events, as well as with events from a single-electron trigger. Photons can belong to four different categories based on their barrel or endcap location and $R_9$ value, and each category has its own preselection criteria based on $H/E$, $\sigma_{i\eta i\eta}$, $R_9(5{\times}5)$, $I_{\PGg}$, and $I_\text{trk}$. The latter variable is a tracker hollow-cone isolation defined as the sum of the transverse momenta of all tracks inside a cone around the photon candidate.
A working point is defined where the BDT output gives a signal efficiency of around 90\% for identifying photons.

\begin{table}[htbp!]
\centering
\topcaption{
Selection criteria applied to simulated $\PGg + \text{jet(s)}$ samples to train the BDT for signal photon identification.
}
\begin{scotch}{lcccccc}
 Region & $R_9$ & $H/E$    & $\sigma_{i\eta i\eta}$    & $R_9 (5{\times}5)$    & $I_{\PGg} (\GeVns{})$    & $I_\text{trk} (\GeVns{})$ \\ \hline
    EB & $>$0.85        & $<$0.08            & \NA                                & $>$0.5            & \NA                & \NA \\ 
    EB & $\leq$0.85     & $<$0.08            & $<$0.015                          & $>$0.5            & $<$4.0            & $<$6.0 \\ 
    EE & $>$0.9         & $<$0.08            & \NA                                & $>$0.8            & \NA                & \NA \\ 
    EE & $\leq$0.9      & $<$0.08            & $<$0.035                          & $>$0.8            & $<$4.0            & $<$6.0 \\
\end{scotch}
\label{tab:ID_preselection}
\end{table}

The MVA ID applies identification and isolation requirements on the reconstructed particle in the ECAL, but does not distinguish between photons and electrons. For this reason, an additional veto is employed to filter out electrons from the photon candidates~\cite{CMS:EGM-14-001}. For the signal samples described in Section~\ref{sec:samples}, the MVA ID is found to be 84.8\% efficient, and the electron veto is found to be 96.2\% efficient for the combination of both signal photons passing the ID.

\section{Event selection}
\label{sec:selection}

For the 2016 data the HLT trigger required a photon candidate with a transverse momentum greater than 60\GeV, and an $H/E$ ratio below 0.15. 
For 2017 and 2018 data, the transverse momentum requirement was raised to 70\GeV.
The efficiency of each of these triggers has been studied as a function of \PT, and selections safely above the trigger threshold (trigger-safe \PT selections) are applied to the data.
For the 2016 data, the trigger-safe \PT selection is placed at 75\GeV and, for the 2017 and 2018 data, the \PT selection is placed at 100\GeV. Additionally, a trigger-safe $H/E$ selection requirement of $H/E<0.10$ is applied.

In addition to the HLT selection, the preselection introduces special criteria to ensure the quality of the diphoton events.
To ensure that the electromagnetic objects selected by the trigger are photons, we use the MVA photon ID and electron veto described in Section~\ref{sec:id}.
Furthermore, the single-photon pseudorapidity is constrained to be within the region $\abs{\eta^{\PGg}} <2.5$, with an additional veto between $1.444 < \abs{\eta^{\PGg}} < 1.566$ to avoid the EE-EB transition region.
Finally, the mass of the diphoton pair must be greater than 350\GeV, a region where the cross section for the SM LbL process is negligible, as motivated in~\cite{Fichet:2014uka}.

In order to reconstruct the exclusive $\PGg\PGg\to\PGg\PGg$ process, photons are selected that are back-to-back in azimuthal angle $\phi$~\cite{dEnterria:2013zqi}, by using the acoplanarity variable defined as
\begin{equation}
    \label{eqn:acoplanarity}
    A_{\phi} \equiv 1 - \abs{\Delta\phi^{\PGg\PGg}/\pi}, \text{ where } \Delta\phi^{\PGg\PGg} = \phi^{\PGg2}-\phi^{\PGg1}.
\end{equation}
Requiring $A_{\phi} < 0.0025$ results in a significant rejection of inclusive diphoton backgrounds.

The final step in the event selection is based on the $\xi_{\PGg\PGg}^+$ and $\xi_{\PGg\PGg}^-$ variables of the diphoton system measured in the central CMS detectors, and defined as 
\begin{equation}
    \label{eqn:cms_xi}
    \xi_{\PGg\PGg}^\pm = \frac{1}{\sqrt{s}}\sum_{i=1}^2 \PT^{\PGg_i}\re^{\pm\eta_{\PGg_i}}.
\end{equation}

In the case of a truly exclusive photon-fusion event, the $\xi$ values of the two intact protons must be compatible with those derived from the central photons as given by Eq.~(\ref{eqn:cms_xi}). Therefore, if events are required to have forward protons within the acceptance of PPS, the diphoton $\xi_{\PGg\PGg}$ values must also be within the PPS $\xi$ range, namely they both are required to satisfy $0.02 < \xi_{\PGg\PGg}^\pm < 0.20$.

Additionally, for the pair of protons detected in the PPS, it was found that placing a lower requirement on the proton $\xi$ values of 0.035 decreases the background (as discussed in Section~\ref{sec:background}) without significantly affecting the signal efficiency.
An asymmetric selection is also placed on the upper end of the proton $\xi$ spectrum to take into account the PPS acceptance.
These asymmetric upper $\xi$ selection criteria are placed at 0.15 for sector-45 and 0.18 for sector-56.
The final selection criteria are listed in Table~\ref{tab:selection_summary}. 

\begin{table*}[htb!]
\centering
\topcaption{Summary of the criteria applied to select exclusive diphoton events.}
\begin{scotch}{ll}
Selection type                  & Criteria \\ \hline
Preselection                    & Diphoton HLT \\
                                & $\pt^{\PGg} > 75\,(100)\GeV$ for 2016\,(2017--2018) \\
                                & $H/E < 0.10$ \\
                                & Photon ID with electron veto \\
                                & $\abs{\eta^{\PGg}} < 2.5$ (except $1.444<\abs{\eta^{\PGg}}<1.566$, EB-EE transition veto)\\
                                & $m_{\PGg\PGg} > 350\GeV$ \\[2ex]
Acoplanarity                    & $A_{\phi} < 0.0025$ \\[2ex]
Diphoton $\xi_{\PGg\PGg}$   & $0.02 < \xi^{\pm}_{\PGg\PGg} < 0.20$ \\[2ex]
Diproton $\xi_{\Pp}$ acceptance & $0.035 < \xi_{\Pp} < 0.150$ (0.180) for sector-45 (sector-56) \\
\end{scotch}
\label{tab:selection_summary}
\end{table*}

\begin{figure}[htpb!]
\centering
\includegraphics[width=\columnwidth]{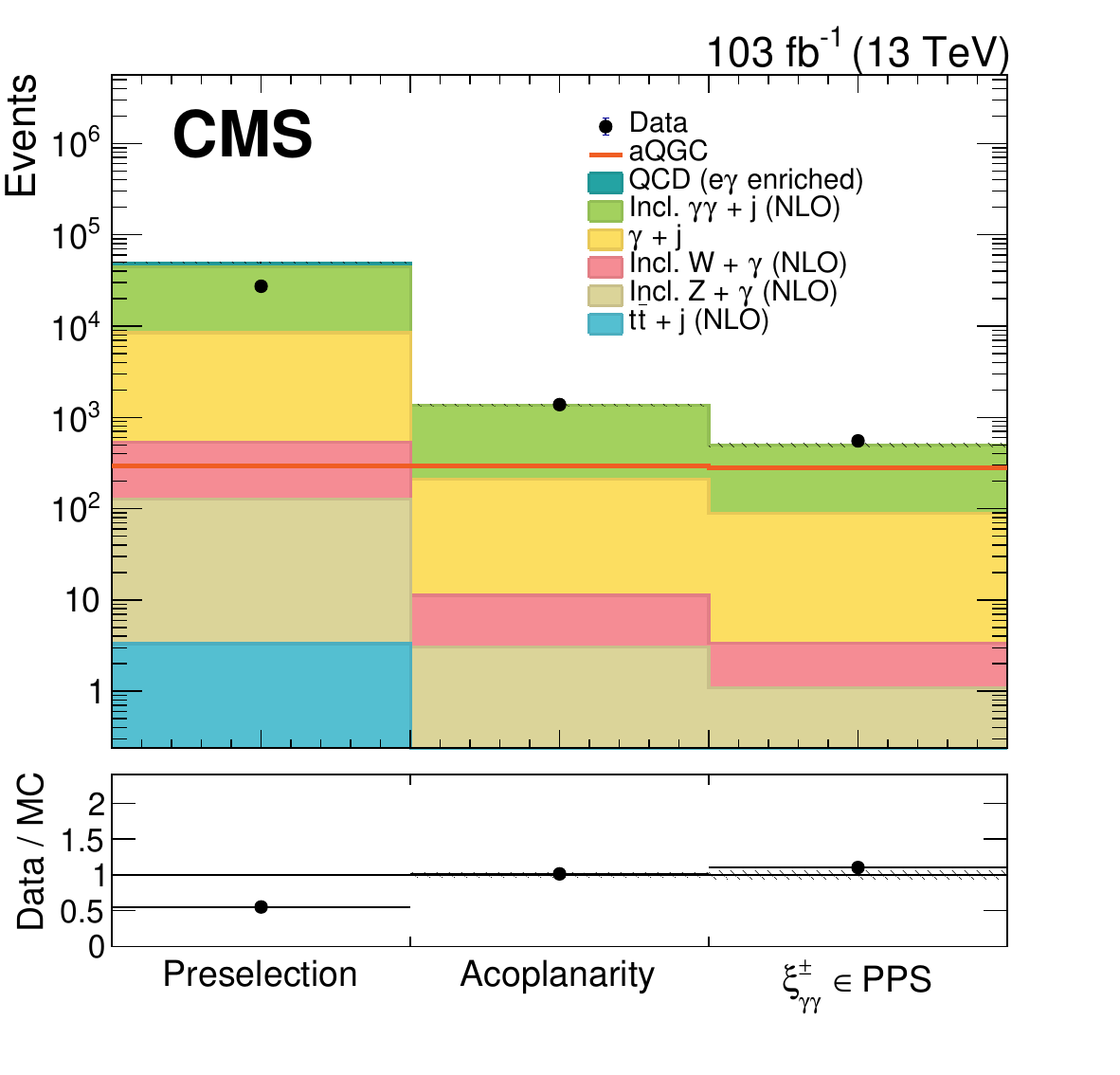}
\caption{Number of events in data (black symbols) and simulated backgrounds (histograms) after each of the three consecutive selection criteria applied (upper plot), and ratio of data to sum of all backgrounds (lower plot). The first bin corresponds to the preselection region, the second bin to the acoplanarity criterion, and the third bin to the diphoton $\xi$ selection defined in the text. The lined red histogram represents an aQGC signal with $\zeta_1 = 0.5\TeV^{-4}$ and $\zeta_2 = 0.1\TeV^{-4}$ for reference. Hatched bands indicate systematic uncertainties (statistical uncertainties are smaller than the symbols size).
}
\label{fig:cutflow}
\end{figure}

\begin{figure*}[htbp!]
    \centering
    \includegraphics[width=0.4\textwidth]{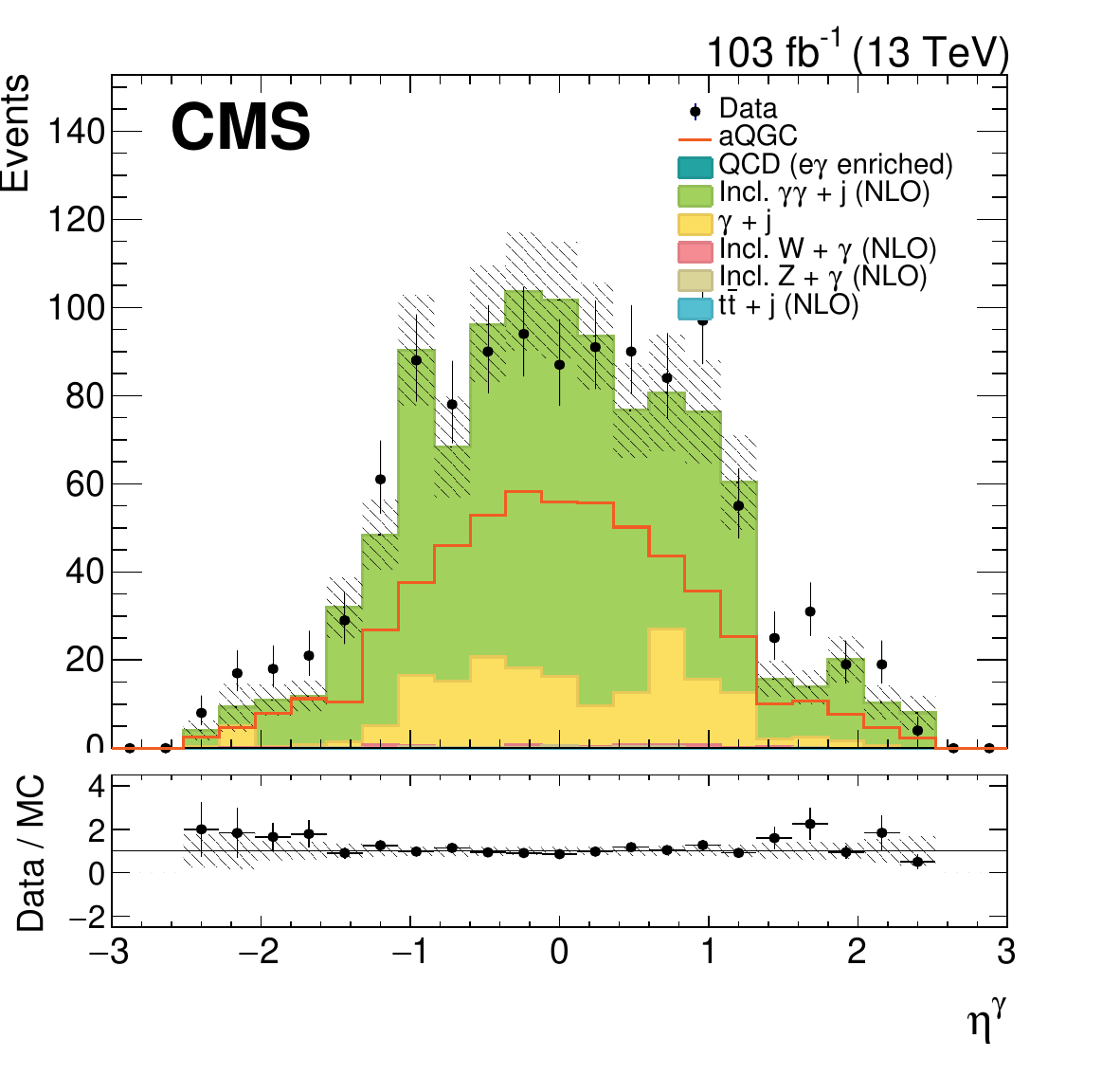}
    \includegraphics[width=0.4\textwidth]{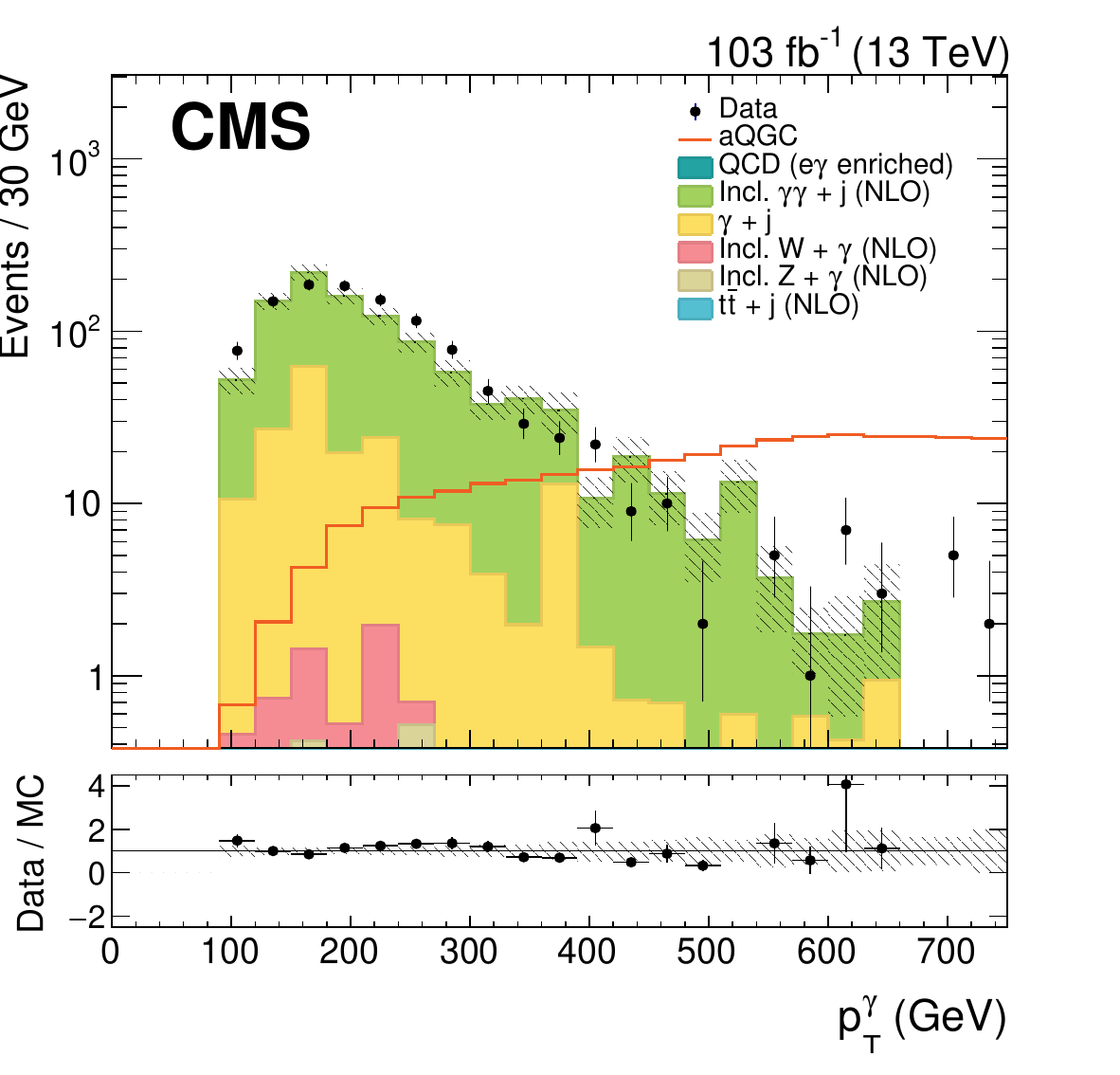}\\
    \includegraphics[width=0.4\textwidth]{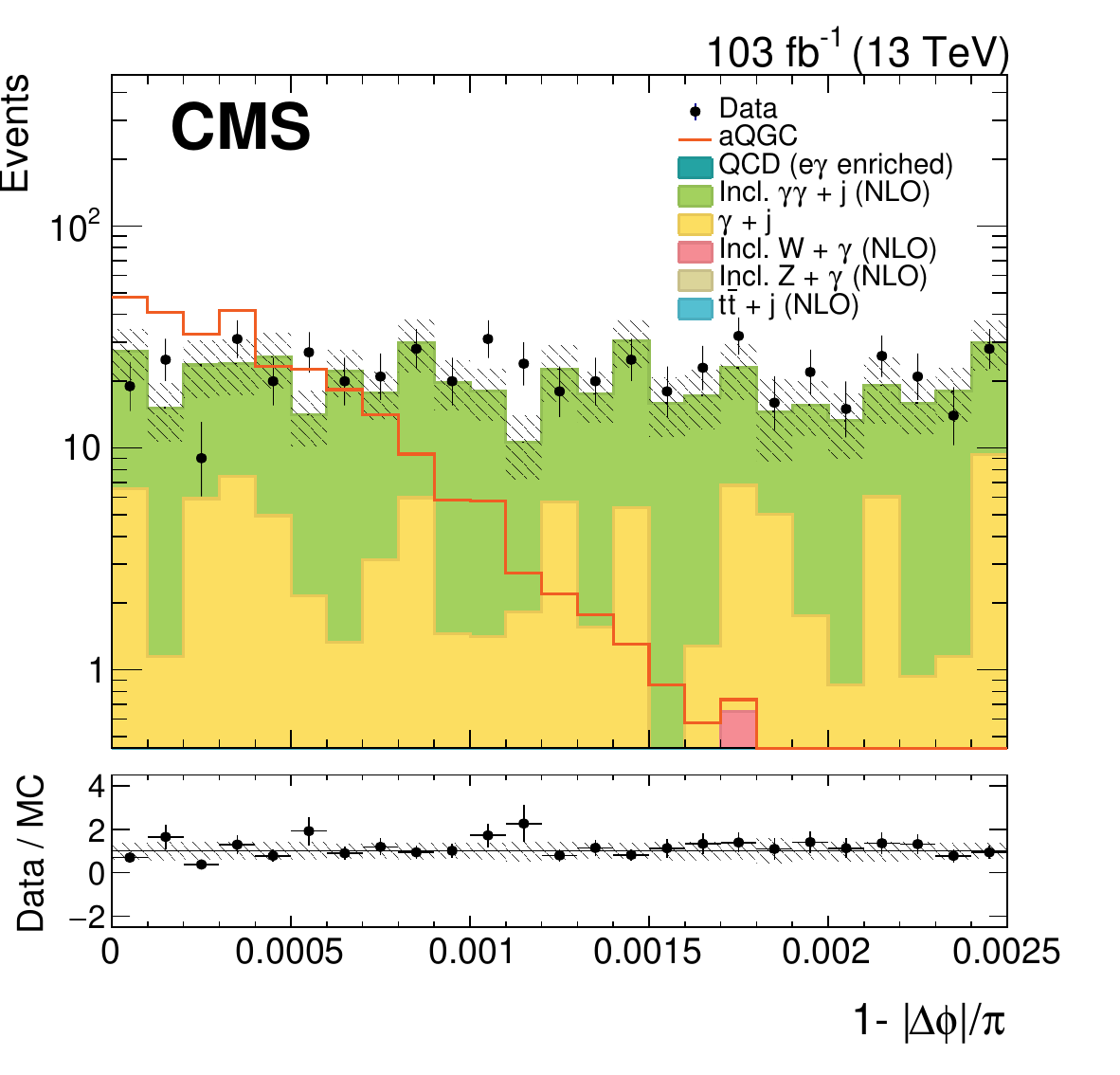}
    \includegraphics[width=0.4\textwidth]{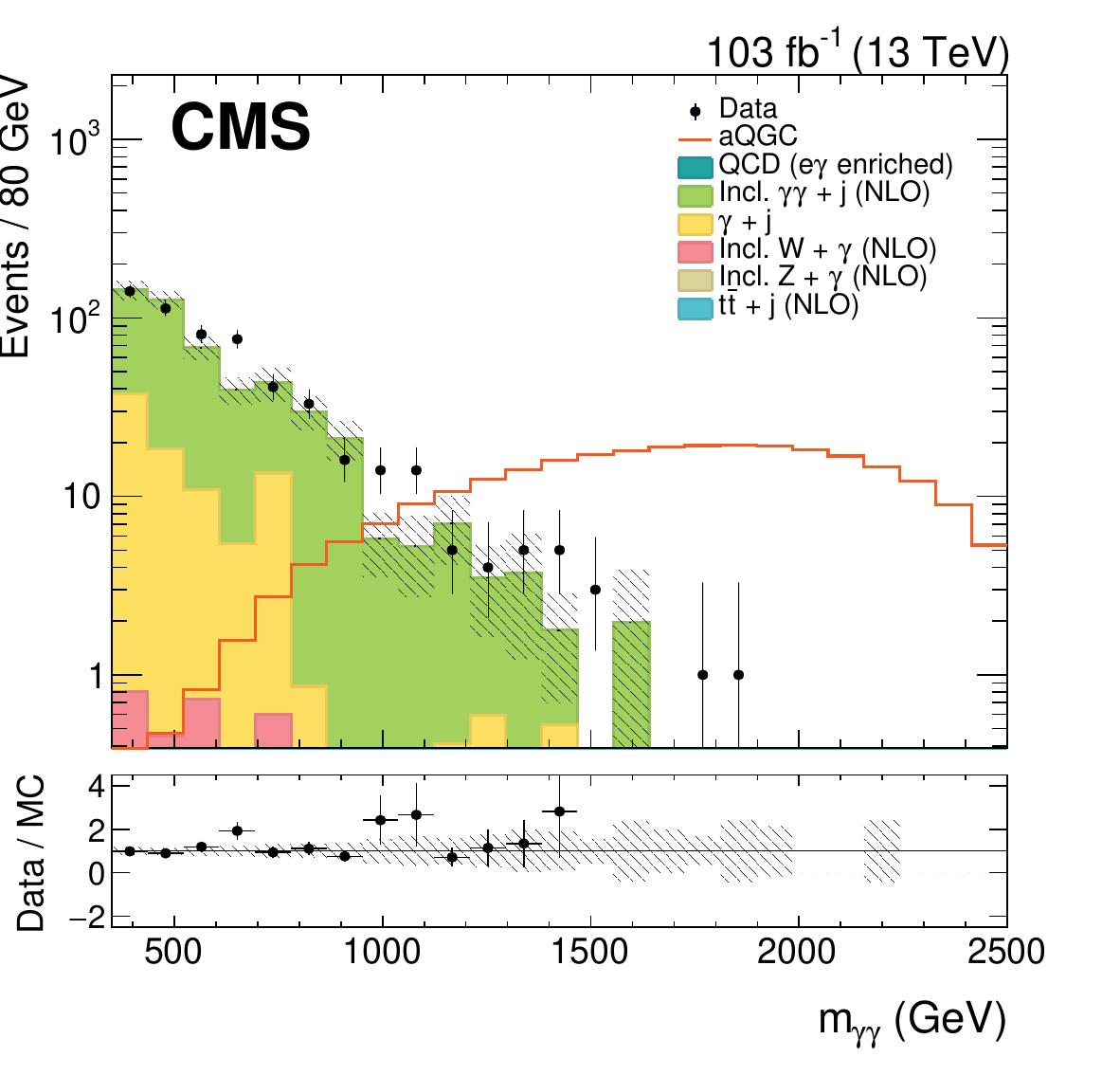}\\
    \includegraphics[width=0.4\textwidth]{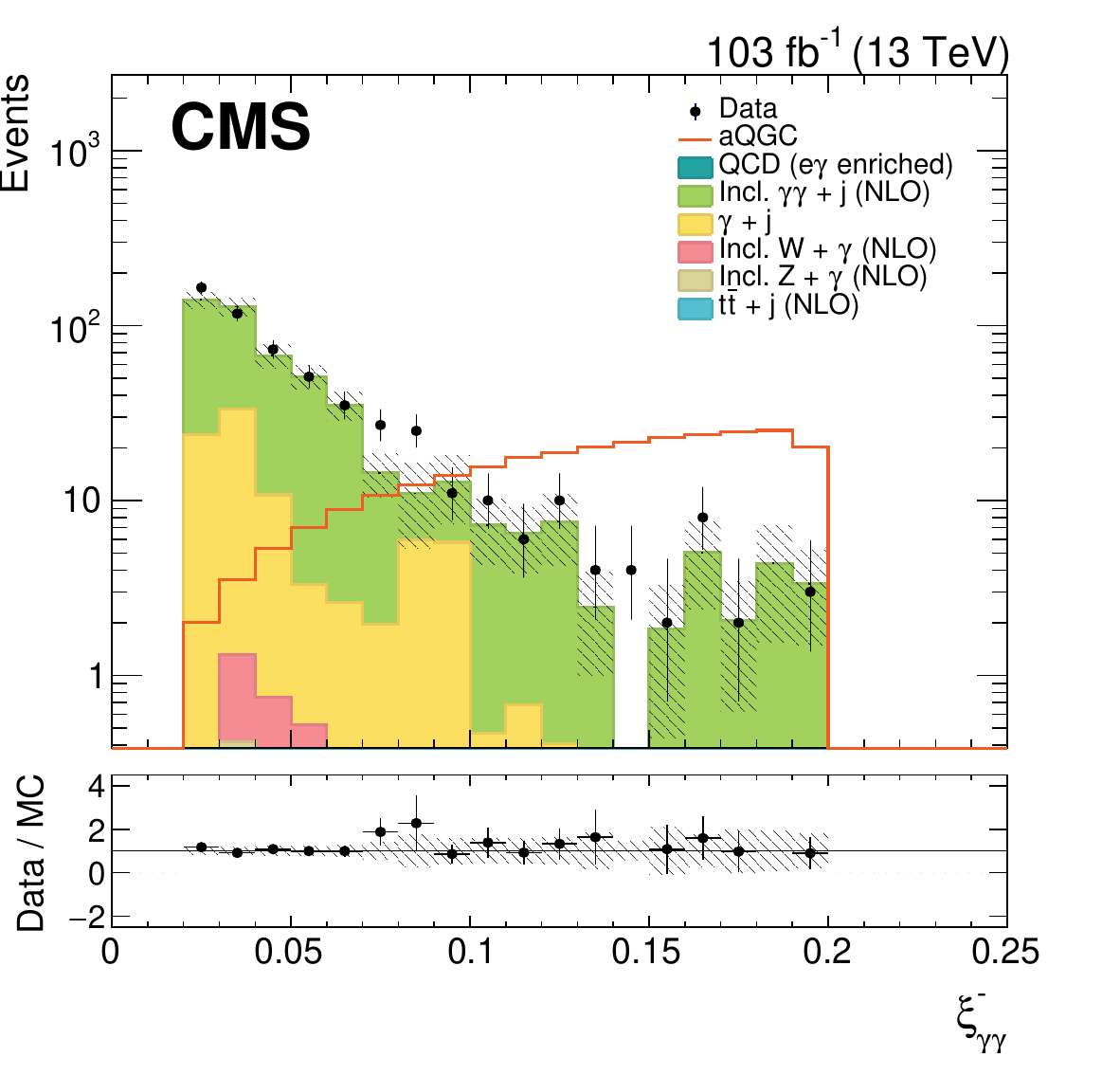}
    \includegraphics[width=0.4\textwidth]{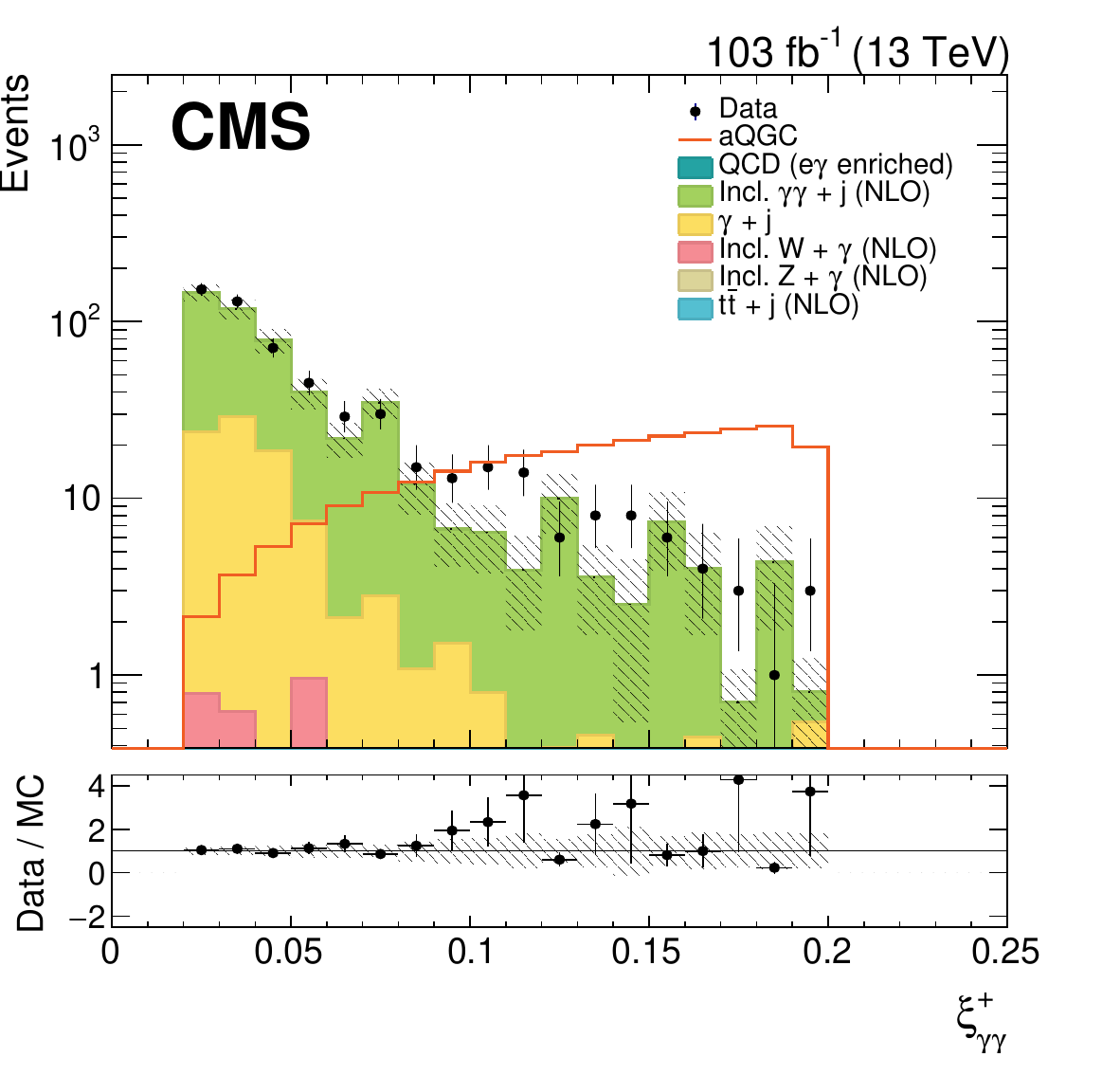}
    \caption{Kinematic distributions for events passing all selection criteria.
    From upper to lower and left to right are the single-photon $\eta$, and \PT, diphoton acoplanarity, mass, $\xi^{-}_{\PGg\PGg}$, and $\xi^{+}_{\PGg\PGg}$ distributions. The black dots represent the data, filled histograms represent the simulated SM backgrounds, and the lined red histogram represents an aQGC signal with $\zeta_1 = 50\TeV^{-4}$ and $\zeta_2 = 100\TeV^{-4}$ for reference. The lower panels in each plot show the ratio of the number of data events to the total SM background expectation. The dashed box indicates the systematic uncertainties.}
    \label{fig:kinematics}
\end{figure*}

Table~\ref{tab:xi_selection} lists the number of simulated events for each background source that pass all selection criteria. Figure~\ref{fig:cutflow} shows the impact of each step in the selection sequence: preselection, acoplanarity selection, and diphoton $\xi$ selection. Good agreement is observed between the MC simulation and data, especially for the number of events after all selection criteria. The kinematic distributions of events passing all selections are shown in Fig.~\ref{fig:kinematics}.

\begin{table}[htpb!]
   \centering
    \topcaption{Summary of the predicted number of events for each SM background contributing to the $\xi \in \text{PPS}$ selection region, their sum, and the observed number of events. The uncertainties quoted are statistical only.}
    \begin{scotch}{lc}
        Sample        & Events\\
        \hline
        Inclusive $\PGg\PGg + \text{jet(s)}$ & $580 \pm 41$\\
        Inclusive $\PGg + \text{jet(s)}$       & $110 \pm 22$\\
        Inclusive $\PW\PGg$        & 4 $\pm2$ \\
        Inclusive $\PZ\PGg$        & 2 $\pm1$ \\
        Inclusive $\ttbar + \text{jet(s)}$      & negligible\\
        QCD jets                   & negligible\\[\cmsTabSkip]
        Total predicted        & $696 \pm 47$ \\
        Observed               & $735 \pm 27$ \\
        Data/MC                & $1.06 \pm 0.08$\\
    \end{scotch}
    \label{tab:xi_selection}
\end{table}

\section{Results}
\label{sec:results}

After all selection criteria (Table~\ref{tab:selection_summary}) are applied, a comparison between the forward diproton and central diphoton kinematic properties is performed.
To reiterate, the analysis strategy is based on a search for an excess above the background expectations for events where the mass and rapidity of the diphoton and diproton systems match.
This is done by comparing the values calculated in Eqs.~(\ref{eqn:massrap}) and (\ref{eqn:cms_xi}) taking into account their associated uncertainties given by
\begin{equation}
    \label{eqn:unc_massrap}
    \frac{\delta m_{\Pp\Pp}}{m_{\Pp\Pp}} = \delta y_{\Pp\Pp} = \frac{1}{2}\left(\frac{\delta\xi^+_\Pp}{\xi^+_\Pp}\oplus \frac{\delta\xi^-_\Pp}{\xi^-_\Pp}\right)
\end{equation}
where $\delta m_{\Pp\Pp}$, $\delta y_{\Pp\Pp}$, and $\delta\xi^\pm_\Pp$ represent the uncertainties on the diproton mass and rapidity, and the uncertainty on the positive and negative proton $\xi$, respectively. 

An event is considered matched if the mass and rapidity of the forward and central systems are within two standard deviations of equivalence. Such a relatively loose matching window is chosen so as to keep as many signal events as possible (given the very low number of background events expected).
The criteria for this matching are:
\begin{equation*}
    \label{eqn:m_2sig}
    \begin{aligned}
    -2 &< \left(m_{\Pp\Pp} - m_{\PGg\PGg}\right) / \delta \left(m_{\Pp\Pp} - m_{\PGg\PGg}\right) &< 2\\
    -2 &< \left(y_{\Pp\Pp} - y_{\PGg\PGg}\right) / \delta \left(y_{\Pp\Pp} - y_{\PGg\PGg}\right) &< 2
    \end{aligned}
\end{equation*}
where $\delta \left(m_{\Pp\Pp} - m_{\PGg\PGg}\right)$ and $\delta \left(y_{\Pp\Pp} - y_{\PGg\PGg}\right)$ are the uncertainties in the differences between the diphoton and diproton mass and rapidity, respectively.
The uncertainties are clearly dominated by the diproton system, and derived event-by-event.

\begin{figure}[htpb!]
    \centering
    \includegraphics[width=\columnwidth]{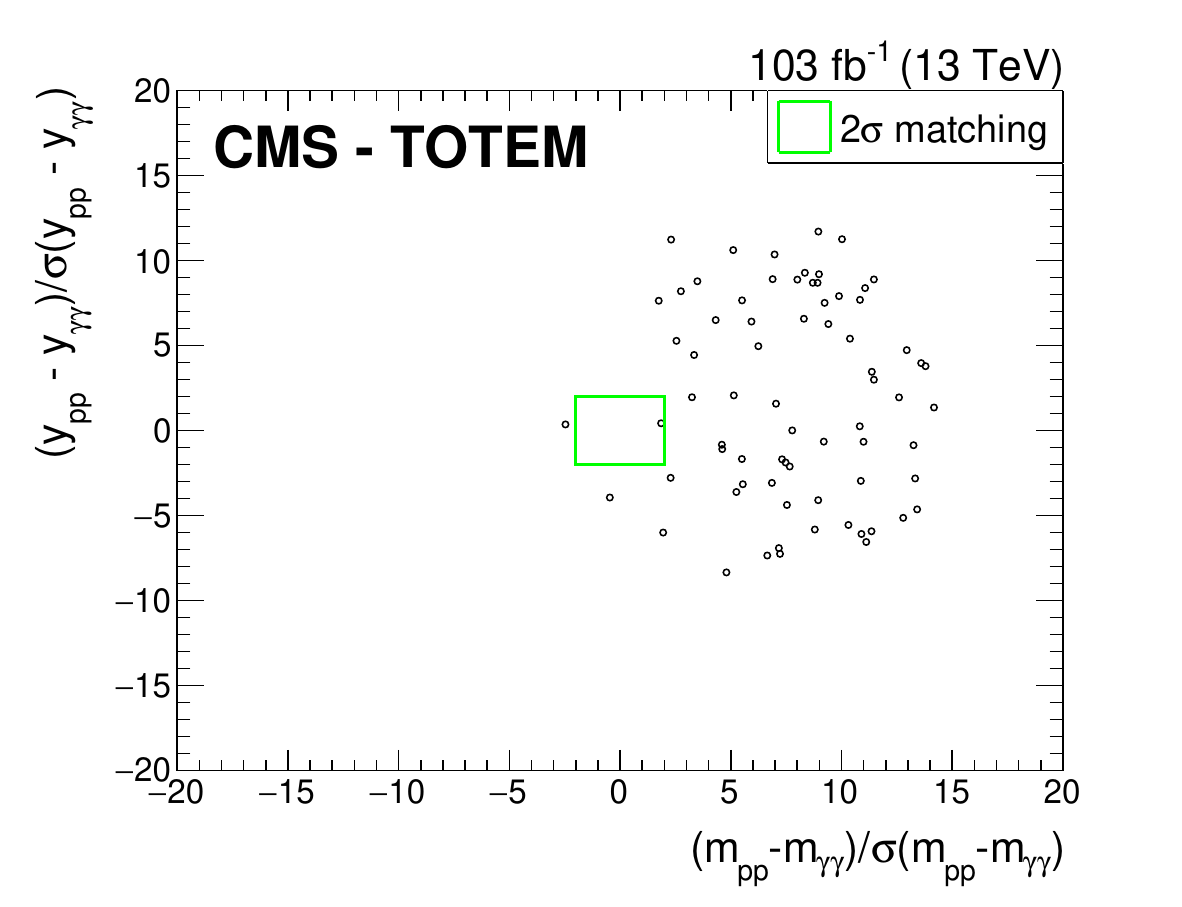}
    \caption{Mass versus rapidity matching distributions for events passing the diphoton and diproton selection criteria described in the text. The matching window is shown by a green rectangle corresponding to two standard deviations.
    }
    \label{fig:matching_2d}
\end{figure}

Figure~\ref{fig:matching_2d} shows events from data as a function of how closely they match (in terms of standard deviations) in mass and rapidity.
One event is observed within the matching window.

\subsection{Background estimation}
\label{sec:background}

The signal region after forward-central matching is expected to have contamination from pileup events. 
The overlap of events with pileup protons detected in PPS and events with inclusively produced diphotons in the central CMS detectors can lead to accidental forward-central matchings. Such events are the only significant source of background for this analysis.

To quantify how these events contaminate the signal region, we use a fully data-driven approach to generate background-like pseudoevents.
As inputs for the creation of the pseudoevents, diphoton events in the signal region are used as well as protons (detected in PPS) from all CMS events passing the diphoton trigger.
Each pseudoevent is created by combining a diphoton pair with a diproton pair randomly added from the same run and LHC crossing-angle.
We then check for the forward-central matching in mass and rapidity.
This procedure is done for all diphoton events in the signal region to create one pseudoexperiment.
Since the protons are randomly mixed with the diphotons, any matching that occurs is truly accidental and not  due to a true correlation between the forward and central systems.
As expected, the number of matching events for the pseudoexperiments follows a Poisson distribution.
The background estimation is validated by using simulated diphoton events from the signal selection as an input to the pseudoexperiment.
A second validation method uses diphoton events from an orthogonal selection (reversing the acoplanarity selection) as an input to the pseudoexperiments.
Both validation methods give similar numbers of average matching events and are also used to extract the systematic uncertainty in the background estimation procedure.

The number of expected matching background events is 0.03, 0.16, and 0.91 for the 2016, 2017, and 2018 data taking, respectively. The total number of expected background events is $1.10 \pm 0.01$~(statistical uncertainty).
The differences in the background estimations between the years are driven by their different luminosities~\cite{CMS:2021xjt,CMS:LUM-17-004,CMS:LUM-18-002}, detector configurations, and efficiencies~\cite{TOTEM:2022vox}.

\subsection{Systematic uncertainties}

The integrated luminosities of the 2016, 2017, and 2018 data-taking periods are individually known with uncertainties in the 1.2--2.5\% range~\cite{CMS:2021xjt,CMS:LUM-17-004,CMS:LUM-18-002}, while the total integrated luminosity has an uncertainty of 1.9\%, with an improvement in precision reflecting the (uncorrelated) time evolution of some systematic effects.
A 23.3, 25.2, and a 20.9\% systematic uncertainty in the background estimation procedure is assigned for 2016, 2017, and 2018, respectively, calculated as the largest relative difference in the number of estimated events between the default method and the two validation methods for the pseudoexperiments.
For the signal efficiency in 2016, 2017, and 2018, uncertainties of 3.1, 7.0, and 2.9\% are assigned, respectively, as derived from the photon ID and assuming that the ID efficiencies are uncorrelated.
A 10\% uncertainty is assigned to the theoretical proton survival probability, \ie, the probability that the interacting protons do not break up due to soft exchanges between their spectator partons, as estimated in~\cite{Harland-Lang:2015cta} and considered in~\cite{CMS:2022dmc}. A 1.7\% uncertainty is assumed for the particle shower reconstruction efficiency in the PPS detectors in 2018~\cite{TOTEM:2022vox}.
The systematic uncertainties of the proton reconstruction are assigned on an event-by-event basis.
The systematic uncertainties described above are summarized in Table~\ref{tab:systematics}.
Systematic uncertainties are treated as uncorrelated across the years.  In the exclusion limit setting procedure, all systematic uncertainties apply to the signal (except the uncertainty associated with the background estimation procedure) and background (except the uncertainty associated with the proton survival probability).

\begin{table}[htpb!]
\centering
\topcaption{Systematic uncertainties corresponding to each year of data taking used in the analysis.}
\begin{scotch}{lccc}
Source                                  & 2016          & 2017          & 2018 \\ \hline
Integrated luminosity                          & 1.2\%         & 2.3\%         & 2.5\% \\
Background estimation                   & 23.3\%        & 25.2\%        & 20.9\% \\
Photon ID scale factors                 & 3.1\%         & 7.0\%         & 2.9\% \\
Proton survival probability       & 10\%          & 10\%          & 10\% \\
Particle shower reconstruction in PPS                 & \NA            & \NA           & 1.7\% \\
\end{scotch}
\label{tab:systematics}
\end{table}

\subsection{Limit on anomalous quartic gauge couplings}

With one event observed and $1.10 \pm 0.24~(\text{stat} + \text{syst})$ events expected, no excess is observed above the SM prediction. Limits are thereby placed on the anomalous coupling parameters, $\zeta_1$ and $\zeta_2$.

The modified frequentist criterion $\CL_\text{s}$~\cite{Junk:1999kv,Read:2002hq} with the profile likelihood ratio test statistic modified for upper limits~\cite{CMS-NOTE-2011-005} and determined by pseudoexperiments is used to evaluate the observed and expected limits at 95\% confidence level (\CL) on the production cross section of aQGC LbL scattering within the fiducial region of the analysis defined as $\pt^{\PGg} > 100\GeV$, $\abs{\eta^{\PGg}} < 2.5$, $m_{\PGg\PGg} > 350\GeV$, and fractional proton energy loss of $0.035 < \xi < 0.150$ (0.180) for the positive-$z$ (negative-$z$) arm of PPS.
The signal efficiency is evaluated over a wide range of the coupling parameters $\zeta_1$ and $\zeta_2$ using FPMC~\cite{Boonekamp:2011ky} and found approximately constant for each year in the search region. 

\begin{figure*}[htpb!]
    \centering
    \includegraphics[width=0.48\textwidth]{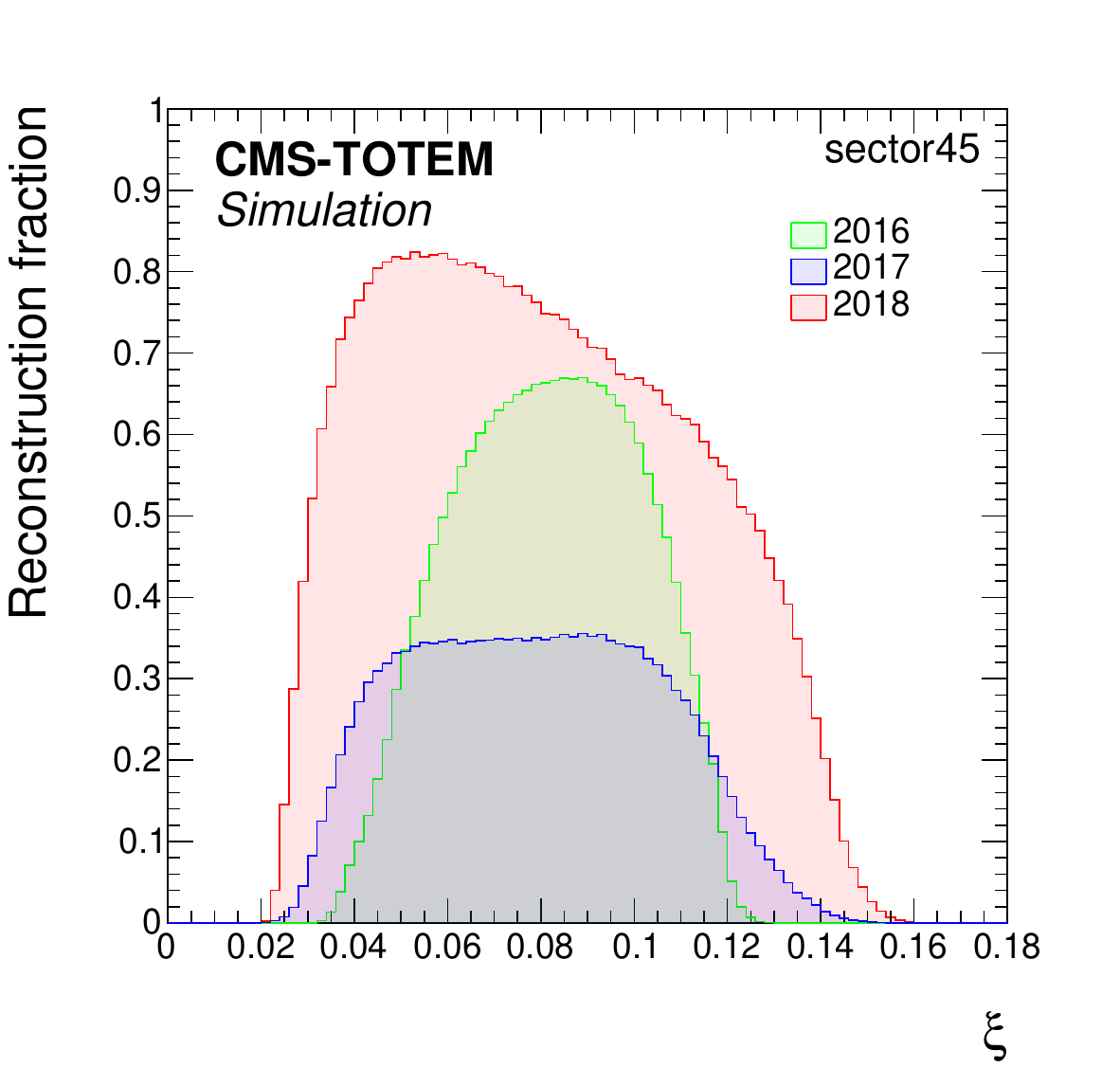}
    \includegraphics[width=0.48\textwidth]{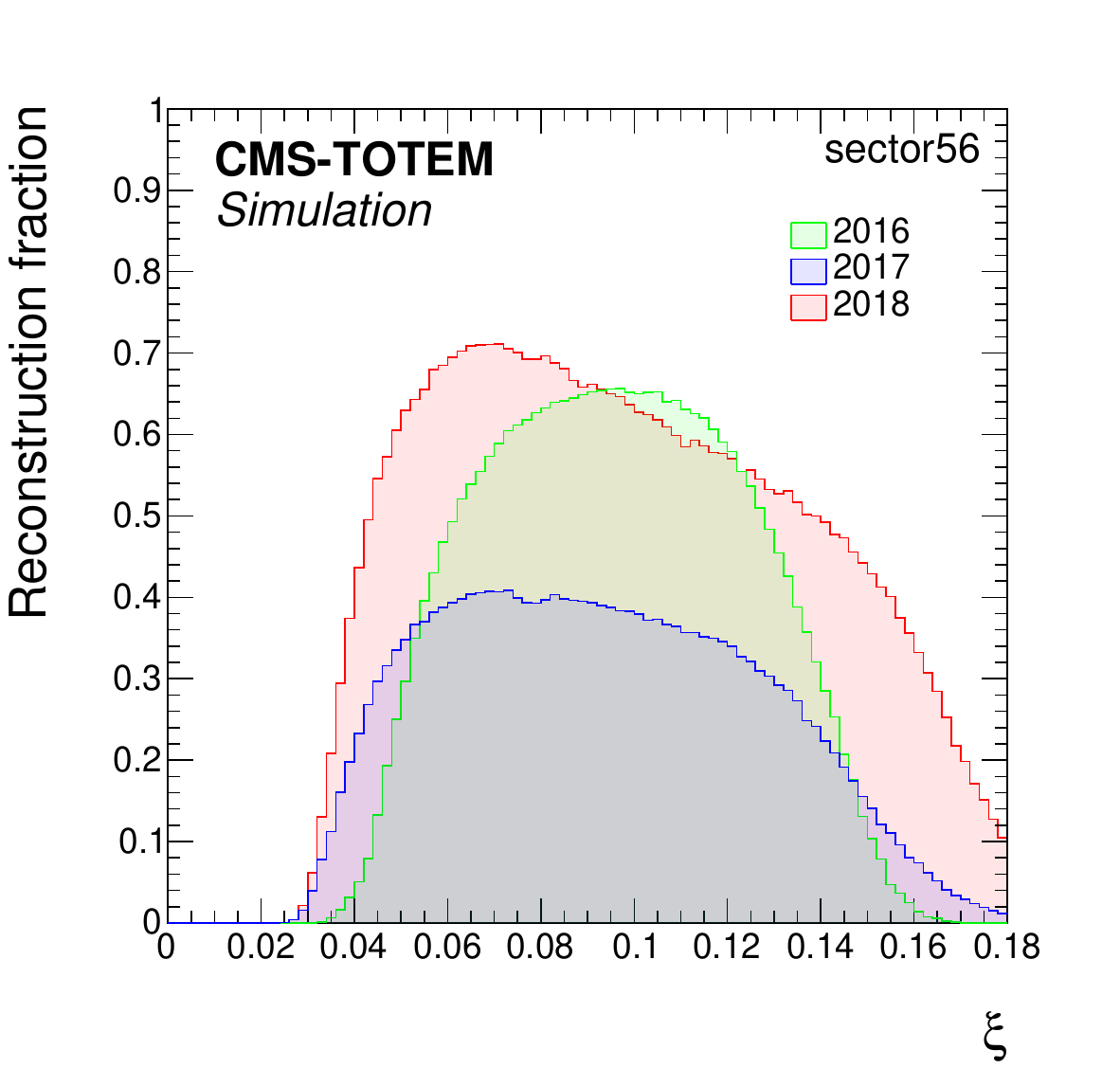}
    \caption{Distributions of combined efficiency-times-acceptance ($\epsilon A$) for diffractively scattered protons as a function of the proton $\xi$ for each running year for PPS sectors~45 (left) and~56 (right), respectively. Differences in the overall reconstruction efficiency are explained by varying detector location, configuration, and design across the years.
    }
    \label{fig:sim_exA}
\end{figure*}

The product of the efficiency ($\epsilon$) and the acceptance ($A$) for the aQGC signal is used as an input to the limit setting tool. The $\epsilon A$ distributions for the PPS detectors to reconstruct diffractively scattered protons as a function of the proton $\xi$ areas shown in Fig.~\ref{fig:sim_exA} for each running year.
The $\epsilon A$ values are calculated as the ratio of the number of reconstructed protons over that of generated protons using a simulation of the forward proton detectors. This simulation accounts for the LHC beam parameters, detector configuration, and detector technology used over the full data-taking period.
The width of the 2016 reconstruction efficiency distribution is narrower than for the other years because of a different physical location of one of the two RPs used for tracking, limiting the overlapping $\xi$ acceptance of the two tracking RPs in this year.
The $\epsilon A$ values of 2017 are overall lower than those of the other years because of the inability of the silicon strip detectors to reconstruct multiple protons within one bunch crossing.
This inefficiency was intensified in 2017 as compared to 2016 because of the increase in the instantaneous luminosity of the LHC.
The signal efficiency is highest in 2018 because of the use of pixel detector technology in all RPs, which do not suffer from the inefficiency of tracking multiple protons.

\begin{figure}[htpb!]
    \centering
    \includegraphics[width=\columnwidth]{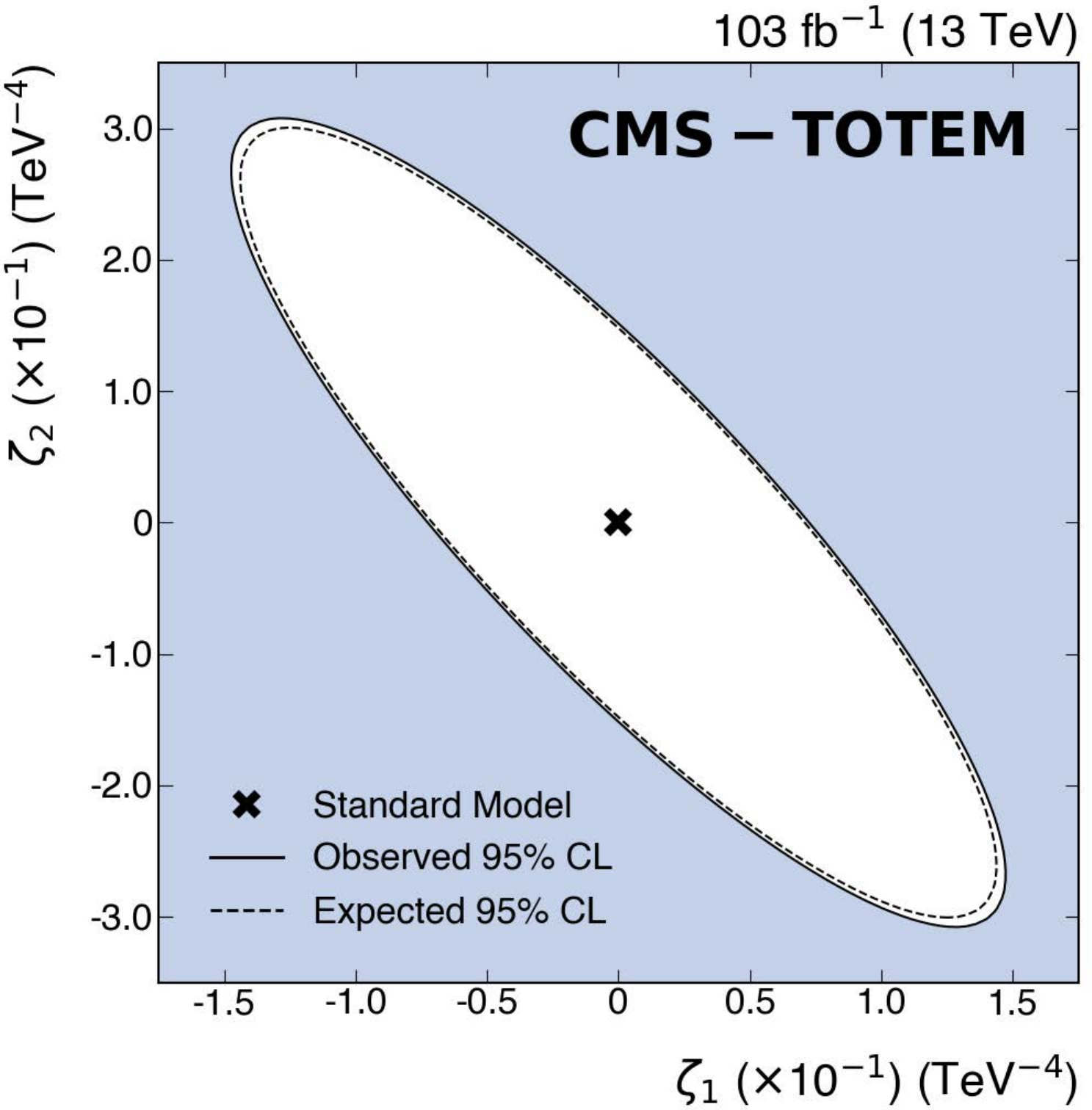}
    \caption{Observed (solid ellipse) and expected (dashed ellipse) exclusion limits at 95\% \CL on the anomalous coupling parameters $\zeta_1$ and $\zeta_2$ derived from the analysis of high-mass exclusive diphoton events in $\Pp\Pp$ collisions at 13 TeV.}
    \label{fig:aqgc_limit}
\end{figure}

The derived $\epsilon A$ values for each year are listed in Table~\ref{tab:exA}.
\begin{table}[htpb!]
\centering
\topcaption{Product of efficiency and acceptance values for each year of data taking, for aQGC signal protons (populating mostly the high-$\xi$ region in Fig.~\ref{fig:sim_exA}). The second column corresponds to the central CMS detectors, the third column to the PPS, and the fourth column is the product of the two previous ones.}
\begin{scotch}{cccc}
Year            & Central-CMS $\epsilon A$ & PPS $\epsilon A$   &  Combined $\epsilon A$ \\
\hline
2016            &  80.1\%              & 6.5\%                        & 5.2\%\\ 
2017            &  75.7\%              & 3.3\%                        & 2.5\%\\ 
2018            &  77.4\%              & 18.4\%                       & 14.2\%\\
\end{scotch}
\label{tab:exA}
\end{table}
Systematic uncertainties are included as nuisance parameters with log-normal probability density functions in the likelihood.
The observed and expected 95\% \CL in the plane of ($\zeta_1$, $\zeta_2$) are shown in Fig.~\ref{fig:aqgc_limit}.

These anomalous $4\PGg$ coupling limits are equivalent to setting an upper limit on the fiducial LbL cross section of
\begin{eqnarray}
  \sigma(\Pp\Pp\to \Pp\PGg\PGg \Pp\mid\xi_{\Pp}\in \xi^\text{PPS})< 0.61\unit{fb},
  \label{eq:upper_limit}
\end{eqnarray}
where $\xi_{\Pp}\in \xi^\text{PPS}$ denotes that both protons are within the acceptance of PPS. This cross section upper limit is the strongest to date in this region of high diphoton masses.

The observed (expected) 95\% \CL limits on the separate anomalous $4\PGg$ coupling parameters (setting the other to zero) are

\begin{equation*}
    \begin{aligned}
    \abs{\zeta_1} < 0.073~(0.071)\TeV^{-4} & (\zeta_2=0),\\
    \abs{\zeta_2} < 0.15~(0.15)\TeV^{-4} & (\zeta_1=0).
    \end{aligned}
\end{equation*}

\subsection{Limit on axion-like particle production}

The same limit setting procedure described above is applied to the $s$-channel production of a pseudoscalar ALP. 
Following the same approach described in~\cite{Baldenegro:2018hng}, this $\PGg\PGg\to a \to\PGg\PGg$ process can be parameterized as a function of the ALP mass, $m_a$, and its photon coupling, $f^{-1}$. 

\begin{figure}[htbp!]
    \centering
    \includegraphics[width=\columnwidth]{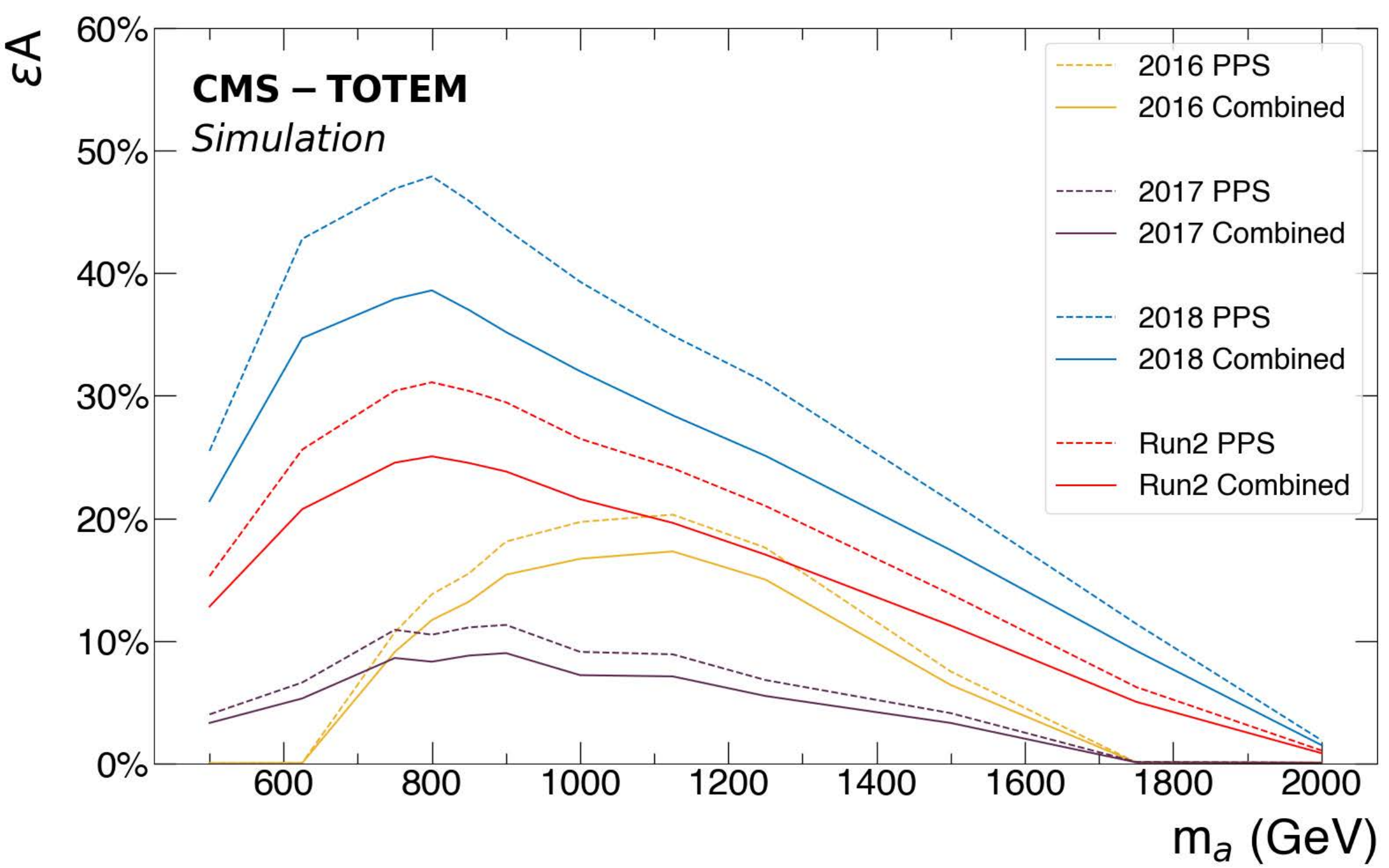}
    \caption{Product of the ALP efficiency and acceptance ($\epsilon A$) values as a function of mass $m_a$ for samples generated with $f^{-1} = 0.1\TeV^{-1}$ ALP-$\PGg$ coupling, within the fiducial volume of this search.
    The curves shown correspond to the PPS $\epsilon A$ values (dashed curves), as well as the latter convolved with the central-CMS $\epsilon A$ (solid curves), for each data-taking year and for the full Run-2.}
    \label{fig:exA_alp}
\end{figure}

For the diphoton invariant mass acceptance determined by the $\xi$ ranges probed in this analysis, the selection efficiency depends only on the mass of the ALP. The $\epsilon A$ product is shown in Fig.~\ref{fig:exA_alp} as a function of ALP mass $m_a$ for a fixed  $f^{-1} = 0.1\TeV^{-1}$ coupling, separately for each year of data taking (as well as for PPS only, and for central-CMS plus PPS detectors) and for the full 2016-2018 data set (Run-2 combined). The distributions of $\epsilon A$ versus ALP mass are used as input to the limit-setting tool. 

\begin{figure}[htpb!]
    \centering
    \includegraphics[width=\columnwidth]{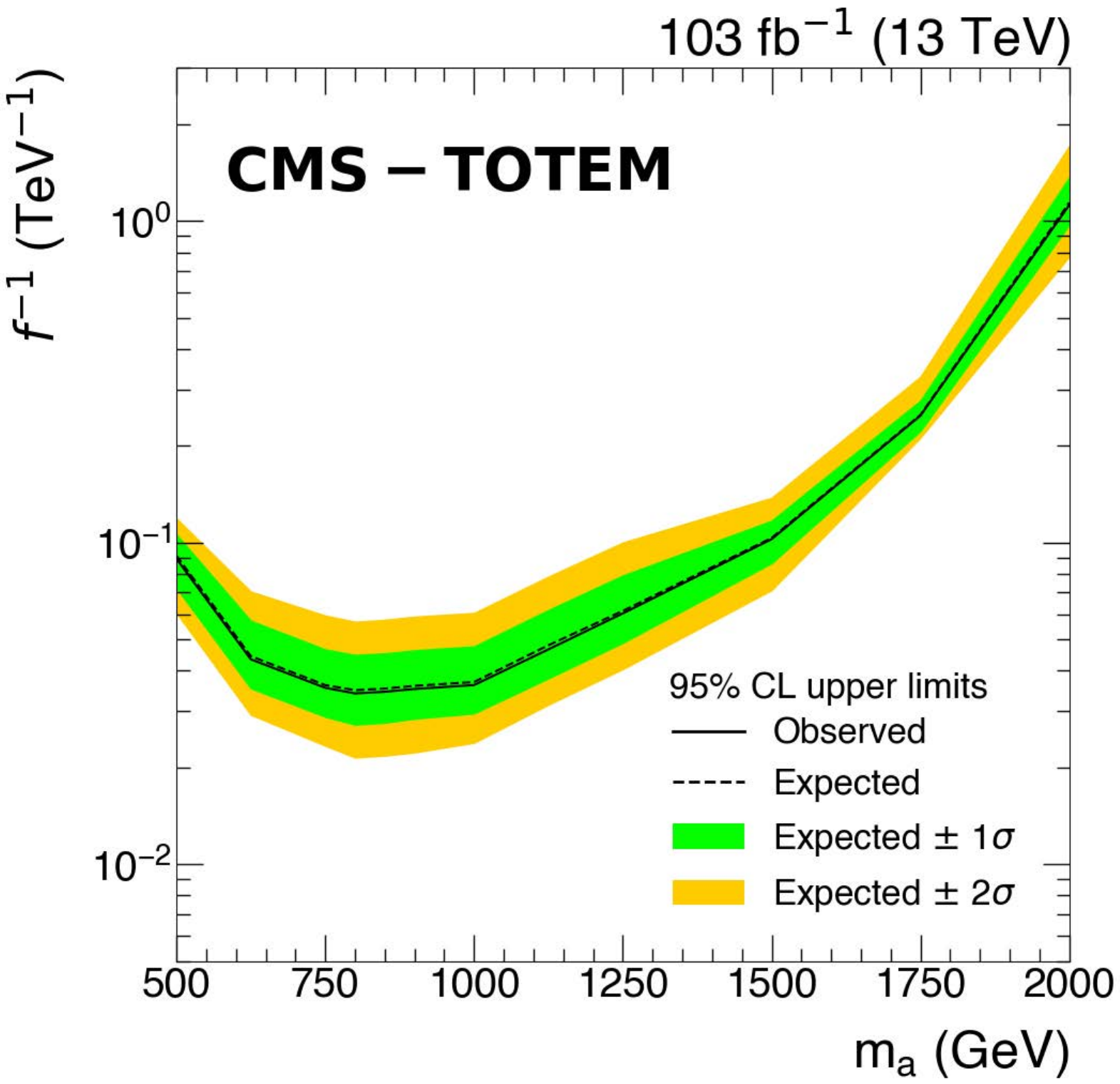}
    \caption{Upper limits at 95\% \CL on ALP-photon coupling strength as a function of the ALP mass.
    The shape of the limit curve is determined by the PPS efficiency-times-acceptance curve.
    The expected limits almost completely overlap with the observed ones.
    }
    \label{fig:alp_limit}
\end{figure}

The observed and expected 95\% \CL limits resulting from the analysis are shown in Fig.~\ref{fig:alp_limit}. Limits on the production of ALPs coupling to photons with strengths $f^{-1} \geq 0.03$ to $1\TeV^{-1}$ are set over $m_a = 500\text{--}2000\GeV$, a region of masses that has been barely explored so far~\cite{dEnterria:2021ljz}. For $m_a \gtrsim 0.7\TeV$, these limits extend those derived in a recent similar analysis by ATLAS (using single-tagged forward protons)~\cite{ATLAS:2023zfc}, and are the strongest to date in this mass range.

\section{Summary}
\label{sec:summary}

A search has been performed for events with a high-mass exclusive diphoton system and two intact protons produced in the final state, $\Pp\Pp\to\Pp\PGg\PGg\Pp$, in proton-proton collisions at $\sqrt{s} = 13\TeV$. The data set corresponds to an integrated luminosity of 103\fbinv collected with the CMS central detectors and the CMS and TOTEM precision proton spectrometer (PPS) in 2016--2018.
Light-by-light (LbL) signal events are selected by requiring the measurement of two high-\PT photons emitted back-to-back, in coincidence with two opposite-side forward protons, measured in PPS, whose kinematic properties match those of the central diphoton system.

The data are found to be in agreement with the predicted standard model (SM) background, with one event observed and $1.10 \pm 0.01 \stat \pm 0.24 \syst$ events expected. An upper limit on the LbL cross section of $\sigma(\Pp\Pp\to \Pp\PGg\PGg \Pp) < 0.61\unit{fb}$ is set within the fiducial range of the analysis defined as $\pt^{\PGg} > 100\GeV$, $\abs{\eta^{\PGg}} < 2.5$, $m_{\PGg\PGg} > 350\GeV$, and fractional proton energy loss of $0.035 < \xi_{\Pp} < 0.150$ (0.180) for the positive-$z$ (negative-$z$) arm of PPS. Limits at 95\% confidence level are derived for the four-photon anomalous quartic gauge couplings (aQGC) parameters (setting, alternatively, the other to zero) $\abs{\zeta_1} < 0.073\TeV^{-4}$ and $\abs{\zeta_2} < 0.15\TeV^{-4}$, using an effective field theory. Additionally, limits on the production of axion-like particles (ALPs) coupling to photons with strengths $f^{-1} \approx 0.03$ to $1\TeV^{-1}$ are set over the mass range from 500 to 2000\GeV. These are the most restrictive limits to date on $4\PGg$ aQGC and on ALPs coupling to photons, in the very high mass phase space region.

\begin{acknowledgments}
We congratulate our colleagues in the CERN accelerator departments for the excellent performance of the LHC and thank the technical and administrative staffs at CERN and at other CMS and TOTEM institutes for their contributions to the success of the common CMS-TOTEM effort. In addition, we gratefully acknowledge the computing centers and personnel of the Worldwide LHC Computing Grid and other centers for delivering so effectively the computing infrastructure essential to our analyses. Finally, we acknowledge the enduring support for the construction and operation of the LHC, the CMS and TOTEM detectors, and the supporting computing infrastructure provided by the following funding agencies: SC (Armenia), BMBWF and FWF (Austria); FNRS and FWO (Belgium); CNPq, CAPES, FAPERJ, FAPERGS, and FAPESP (Brazil); MES and BNSF (Bulgaria); CERN; CAS, MoST, and NSFC (China); MINCIENCIAS (Colombia); MSES and CSF (Croatia); RIF (Cyprus); SENESCYT (Ecuador); MoER, ERC PUT and ERDF (Estonia); Academy of Finland, Magnus Ehrnrooth Foundation, MEC, HIP, and Waldemar von Frenckell Foundation (Finland); CEA and CNRS/IN2P3 (France); SRNSF (Georgia); BMBF, DFG, and HGF (Germany); GSRI (Greece); NKFIH (Hungary); DAE and DST (India); IPM (Iran); SFI (Ireland); INFN (Italy); MSIP and NRF (Republic of Korea); MES (Latvia); LAS (Lithuania); MOE and UM (Malaysia); BUAP, CINVESTAV, CONACYT, LNS, SEP, and UASLP-FAI (Mexico); MOS (Montenegro); MBIE (New Zealand); PAEC (Pakistan); MES and NSC (Poland); FCT (Portugal); MESTD (Serbia); MCIN/AEI and PCTI (Spain); MOSTR (Sri Lanka); Swiss Funding Agencies (Switzerland); MST (Taipei); MHESI and NSTDA (Thailand); TUBITAK and TENMAK (Turkey); NASU (Ukraine); STFC (United Kingdom); DOE and NSF (USA).

\hyphenation{Rachada-pisek} Individuals have received support from the Marie-Curie program and the European Research Council and Horizon 2020 Grant, contract Nos.\ 675440, 724704, 752730, 758316, 765710, 824093, and COST Action CA16108 (European Union); the Leventis Foundation; the Alfred P.\ Sloan Foundation; the Alexander von Humboldt Foundation; the Science Committee, project no. 22rl-037 (Armenia); the Belgian Federal Science Policy Office; the Fonds pour la Formation \`a la Recherche dans l'Industrie et dans l'Agriculture (FRIA-Belgium); the Agentschap voor Innovatie door Wetenschap en Technologie (IWT-Belgium); the F.R.S.-FNRS and FWO (Belgium) under the ``Excellence of Science -- EOS" -- be.h project n.\ 30820817; the Beijing Municipal Science \& Technology Commission, No. Z191100007219010 and Fundamental Research Funds for the Central Universities (China); the Ministry of Education, Youth and Sports (MEYS) of the Czech Republic; Svenska Kulturfonden (Finland); the Shota Rustaveli National Science Foundation, grant FR-22-985 (Georgia); the Deutsche Forschungsgemeinschaft (DFG), under Germany's Excellence Strategy -- EXC 2121 ``Quantum Universe" -- 390833306, and under project number 400140256 - GRK2497; the Hellenic Foundation for Research and Innovation (HFRI), Project Number 2288 (Greece); the Hungarian Academy of Sciences, the New National Excellence Program - \'UNKP, the NKFIH research grants K 124845, K 124850, K 128713, K 128786, K 129058, K 131991, K 133046, K 138136, K 143460, K 143477, 2020-2.2.1-ED-2021-00181, and TKP2021-NKTA-64 (Hungary); the Council of Science and Industrial Research, India; ICSC -- National Research Center for High Performance Computing, Big Data and Quantum Computing, funded by the EU NexGeneration program (Italy); the Latvian Council of Science; the Ministry of Education and Science, project no. 2022/WK/14, and the National Science Center, contracts Opus 2021/41/B/ST2/01369 and 2021/43/B/ST2/01552 (Poland); the Funda\c{c}\~ao para a Ci\^encia e a Tecnologia, grant CEECIND/01334/2018 (Portugal); the National Priorities Research Program by Qatar National Research Fund; MCIN/AEI/10.13039/501100011033, ERDF ``a way of making Europe", and the Programa Estatal de Fomento de la Investigaci{\'o}n Cient{\'i}fica y T{\'e}cnica de Excelencia Mar\'{\i}a de Maeztu, grant MDM-2017-0765 and Programa Severo Ochoa del Principado de Asturias (Spain); the Chulalongkorn Academic into Its 2nd Century Project Advancement Project, and the National Science, Research and Innovation Fund via the Program Management Unit for Human Resources \& Institutional Development, Research and Innovation, grant B05F650021 (Thailand); the Kavli Foundation; the Nvidia Corporation; the SuperMicro Corporation; the Welch Foundation, contract C-1845; and the Weston Havens Foundation (USA).
\end{acknowledgments}

\bibliography{auto_generated}
\cleardoublepage \appendix\section{The CMS Collaboration \label{app:collab}}\begin{sloppypar}\hyphenpenalty=5000\widowpenalty=500\clubpenalty=5000\input{EXO-21-007-public-authorlist.tex}\end{sloppypar}
\end{document}

%% file: EXO-21-007-public-authorlist.tex
\cmsinstitute{Yerevan Physics Institute, Yerevan, Armenia}
{\tolerance=6000
A.~Tumasyan\cmsorcid{0009-0000-0684-6742}
\par}
\cmsinstitute{Institut f\"{u}r Hochenergiephysik, Vienna, Austria}
{\tolerance=6000
W.~Adam\cmsorcid{0000-0001-9099-4341}, J.W.~Andrejkovic, T.~Bergauer\cmsorcid{0000-0002-5786-0293}, S.~Chatterjee\cmsorcid{0000-0003-2660-0349}, K.~Damanakis\cmsorcid{0000-0001-5389-2872}, M.~Dragicevic\cmsorcid{0000-0003-1967-6783}, A.~Escalante~Del~Valle\cmsorcid{0000-0002-9702-6359}, P.S.~Hussain\cmsorcid{0000-0002-4825-5278}, M.~Jeitler\cmsAuthorMark{1}\cmsorcid{0000-0002-5141-9560}, N.~Krammer\cmsorcid{0000-0002-0548-0985}, L.~Lechner\cmsorcid{0000-0002-3065-1141}, D.~Liko\cmsorcid{0000-0002-3380-473X}, I.~Mikulec\cmsorcid{0000-0003-0385-2746}, P.~Paulitsch, F.M.~Pitters, J.~Schieck\cmsAuthorMark{1}\cmsorcid{0000-0002-1058-8093}, R.~Sch\"{o}fbeck\cmsorcid{0000-0002-2332-8784}, D.~Schwarz\cmsorcid{0000-0002-3821-7331}, M.~Sonawane\cmsorcid{0000-0003-0510-7010}, S.~Templ\cmsorcid{0000-0003-3137-5692}, W.~Waltenberger\cmsorcid{0000-0002-6215-7228}, C.-E.~Wulz\cmsAuthorMark{1}\cmsorcid{0000-0001-9226-5812}
% ...
\par}
\vskip\cmsinstskip
\dag:~Deceased\\
$^{1}$Also at TU Wien, Vienna, Austria\\
% ...

\section{The TOTEM Collaboration\label{app:totem}}
\newcommand{\AddAuthor}[2]{#1$^{#2}$,\ }
\newcommand\AddAuthorLast[2]{#1$^{#2}$}
\noindent
\AddAuthor{G.~Antchev\cmsorcid{0000-0003-3210-5037}}{a}
\AddAuthor{P.~Aspell}{8}
\AddAuthor{I.~Atanassov\cmsorcid{0000-0002-5728-9103}}{a}
\AddAuthor{V.~Avati}{7,8}
\AddAuthor{J.~Baechler}{8}
\AddAuthor{C.~Baldenegro~Barrera\cmsorcid{0000-0002-6033-8885}}{11}
\AddAuthor{V.~Berardi\cmsorcid{0000-0002-8387-4568}}{4a,4b}
\AddAuthor{M.~Berretti\cmsorcid{0000-0003-4122-8282}}{2a}
\AddAuthor{V.~Borshch\cmsorcid{0000-0002-5479-1982}}{12}
\AddAuthor{E.~Bossini\cmsorcid{0000-0002-2303-2588}}{6a}
\AddAuthor{U.~Bottigli\cmsorcid{0000-0002-0666-3433}}{6b}
\AddAuthor{M.~Bozzo\cmsorcid{0000-0002-1715-0457}}{5a,5b}
\AddAuthor{H.~Burkhardt}{8}
\AddAuthor{F.S.~Cafagna\cmsorcid{0000-0002-7450-4784}}{4a}
\AddAuthor{M.G.~Catanesi\cmsorcid{0000-0002-2987-7688}}{4a}
\AddAuthor{M.~Deile\cmsorcid{0000-0001-5085-7270}}{8}
\AddAuthor{F.~De~Leonardis}{4a,4c}
\AddAuthor{M.~Doubek\cmsorcid{0000-0002-0026-9558}}{1c}
\AddAuthor{D.~Druzhkin}{3a,8}
\AddAuthor{K.~Eggert}{10}
\AddAuthor{V.~Eremin}{c}
\AddAuthor{A.~Fiergolski}{8}
\AddAuthor{F.~Garcia\cmsorcid{0000-0002-4023-7964}}{2a}
\AddAuthor{V.~Georgiev}{1a}
\AddAuthor{S.~Giani}{8}
\AddAuthor{L.~Grzanka\cmsorcid{0000-0002-3599-854X}}{7}
\AddAuthor{J.~Hammerbauer}{1a}
\AddAuthor{T.~Isidori\cmsorcid{0000-0002-7934-4038}}{11}
\AddAuthor{V.~Ivanchenko\cmsorcid{0000-0002-1844-5433}}{12}
\AddAuthor{M.~Janda\cmsorcid{0000-0001-5736-6183}}{1c}
\AddAuthor{A.~Karev}{8}
\AddAuthor{J.~Ka\v{s}par\cmsorcid{0000-0001-5639-2267}}{1b,8}
\AddAuthor{B.~Kaynak\cmsorcid{0000-0003-3857-2496}}{9}
\AddAuthor{J.~Kopal\cmsorcid{0000-0001-9751-7409}}{8}
\AddAuthor{V.~Kundr\'{a}t\cmsorcid{0000-0003-2868-5550}}{1b}
\AddAuthor{S.~Lami\cmsorcid{0000-0001-9492-0147}}{6a}
\AddAuthor{R.~Linhart}{1a}
\AddAuthor{C.~Lindsey}{11}
\AddAuthor{M.V.~Lokaj\'{i}\v{c}ek\cmsorcid{0000-0002-2052-1220}$^{\textrm{\dag,}}$}{1b}
\AddAuthor{L.~Losurdo\cmsorcid{0000-0002-4964-7951}}{6b}
\AddAuthor{F.~Lucas~Rodr\'{i}guez}{8}
\AddAuthor{M.~Macr\'{i}}{\dag,5a}
\AddAuthor{M.~Malawski\cmsorcid{0000-0001-6005-0243}}{7}
\AddAuthor{N.~Minafra\cmsorcid{0000-0003-4002-1888}}{11}
\AddAuthor{S.~Minutoli}{5a}
\AddAuthor{K.~Misan\cmsorcid{0009-0007-2416-042X}}{7}
\AddAuthor{T.~Naaranoja\cmsorcid{0000-0001-5797-7929}}{2a,2b}
\AddAuthor{F.~Nemes\cmsorcid{0000-0002-1451-6484}}{3a,3b,8}
\AddAuthor{H.~Niewiadomski}{10}
\AddAuthor{E.~Oliveri\cmsorcid{0000-0002-0832-6975}}{8}
\AddAuthor{F.~Oljemark\cmsorcid{0000-0003-0121-2761}}{2a,2b}
\AddAuthor{M.~Oriunno}{b}
\AddAuthor{K.~\"{O}sterberg\cmsorcid{0000-0003-4807-0414}}{2a,2b}
\AddAuthor{S.~Ozkorucuklu\cmsorcid{0000-0001-5153-9266}}{9}
\AddAuthor{P.~Palazzi\cmsorcid{0000-0002-4861-391X}}{8}
\AddAuthor{V.~Passaro\cmsorcid{0000-0003-0802-4464}}{4a,4c}
\AddAuthor{Z.~Peroutka}{1a}
\AddAuthor{O.~Potok\cmsorcid{0009-0005-1141-6401}}{9}
\AddAuthor{J.~Proch\'{a}zka\cmsorcid{0000-0002-2774-2245}}{1b}
\AddAuthor{M.~Quinto\cmsorcid{0000-0002-6363-6132}}{4a,4b}
\AddAuthor{E.~Radermacher}{8}
\AddAuthor{E.~Radicioni\cmsorcid{0000-0002-2231-1067}}{4a}
\AddAuthor{F.~Ravotti\cmsorcid{0000-0002-5709-6934}}{8}
\AddAuthor{C.~Royon\cmsorcid{0000-0002-7672-9709}}{11}
\AddAuthor{G.~Ruggiero}{8}
\AddAuthor{H.~Saarikko}{2a,2b}
\AddAuthor{V.D.~Samoylenko}{c}
\AddAuthor{A.~Scribano\cmsorcid{0000-0002-4338-6332}}{6a,11}
\AddAuthor{J.~\v{S}irok\'{y}}{1a}
\AddAuthor{J.~Smajek}{8}
\AddAuthor{W.~Snoeys\cmsorcid{0000-0003-3541-9066}}{8}
\AddAuthor{R.~Stefanovitch}{3a,8}
\AddAuthor{C.~Taylor\cmsorcid{0000-0001-6816-8051}}{10}
\AddAuthor{E.~Tcherniaev\cmsorcid{0000-0002-3685-0635}}{12}
\AddAuthor{N.~Turini\cmsorcid{0000-0002-9395-5230}}{6b}
\AddAuthor{O.~Urban}{1a}
\AddAuthor{V.~Vacek\cmsorcid{0000-0001-9584-0392}}{1c}
\AddAuthor{O.~Vavroch}{1a}
\AddAuthor{J.~Welti}{2a,2b}
\AddAuthor{J.~Williams\cmsorcid{0000-0002-9810-7097}}{11}
\AddAuthorLast{J.~Zich}{1a}
\vskip 4pt plus 4pt
\let\thefootnote\relax
\newcommand{\AddInstitute}[2]{${}^{#1}$#2\\}
\newcommand{\AddExternalInstitute}[2]{\footnote{${}^{#1}$ #2}}
\noindent
\dag Deceased\\
\AddInstitute{1a}{University of West Bohemia, Pilsen, Czech Republic}
\AddInstitute{1b}{Institute of Physics of the Academy of Sciences of the Czech Republic, Prague, Czech Republic}
\AddInstitute{1c}{Czech Technical University, Prague, Czech Republic}
\AddInstitute{2a}{Helsinki Institute of Physics, University of Helsinki, Helsinki, Finland}
\AddInstitute{2b}{Department of Physics, University of Helsinki, Helsinki, Finland}
\AddInstitute{3a}{Wigner Research Centre for Physics, RMKI, Budapest, Hungary}
\AddInstitute{3b}{MATE Institute of Technology KRC, Gy\"{o}ngy\"{o}s, Hungary}
\AddInstitute{4a}{INFN Sezione di Bari, Bari, Italy}
\AddInstitute{4b}{Dipartimento Interateneo di Fisica di Bari, University of Bari, Bari, Italy}
\AddInstitute{4c}{Dipartimento di Ingegneria Elettrica e dell'Informazione --- Politecnico di Bari, Bari, Italy}
\AddInstitute{5a}{INFN Sezione di Genova, Genova, Italy}
\AddInstitute{5b}{Universit\`{a} degli Studi di Genova, Genova, Italy}
\AddInstitute{6a}{INFN Sezione di Pisa, Pisa, Italy}
\AddInstitute{6b}{Universit\`{a} degli Studi di Siena and Gruppo Collegato INFN di Siena, Siena, Italy}
\AddInstitute{7}{Akademia G\'{o}rniczo-Hutnicza (AGH) University of Krak\'ow, Faculty of Computer Science, Krak\'ow, Poland}
\AddInstitute{8}{CERN, Geneva, Switzerland}
\AddInstitute{9}{Istanbul University, Istanbul, Turkey}
\AddInstitute{10}{Case Western Reserve University, Department of Physics, Cleveland, Ohio, USA}
\AddInstitute{11}{The University of Kansas, Lawrence, Kansas, USA}
\AddInstitute{12}{Authors affiliated with an institute or an international laboratory covered by a cooperation agreement with CERN}
\AddExternalInstitute{a}{INRNE-BAS, Institute for Nuclear Research and Nuclear Energy, Bulgarian Academy of Sciences, Sofia, Bulgaria}
\AddExternalInstitute{b}{SLAC, Stanford University, California, USA}
\AddExternalInstitute{c}{Authors affiliated with an institute or an international laboratory covered by a cooperation agreement with CERN}